%% file: LamKrip_long.tex
\documentclass{LMCS}

\def\doi{7 (3:18) 2011}
\lmcsheading%
{\doi}
{1--25}
{}
{}
{Apr.~\phantom04, 2011}
{Sep.~27, 2011}
{}   

\usepackage{graphicx,url}
\usepackage{amsmath}
\usepackage{amssymb}
\usepackage{stmaryrd}

\usepackage{tikz}
\usetikzlibrary{arrows}

	\newenvironment{theorem}{\begin{thm}}{\end{thm}}
	\newenvironment{lemma}{\begin{lem}}{\end{lem}}
	\newenvironment{definition}{\begin{defi}}{\end{defi}}
	\newenvironment{example}{\begin{exa}}{\end{exa}}
	
	\newcommand{\myqed}{\qed}

	\usepackage{enumerate}
	\renewenvironment{itemize}{\begin{enumerate}[$\bullet$]}{\end{enumerate}}

\usepackage{typetheory}
\usepackage{basics}
\usepackage{ded}

\usepackage{localcommands}

\usepackage{enumerate,hyperref}

\begin{document}

\title{Kripke Semantics for Martin-L\"of's Extensional Type Theory\rsuper*}

\author[S.~Awodey]{Steve Awodey\rsuper a}
\address{{\lsuper a}Carnegie Mellon University, Pittsburgh, USA}
\email{awodey@cmu.edu}

\author[F.~Rabe]{Florian Rabe\rsuper b}
\address{{\lsuper b}Jacobs University Bremen, Germany}
\email{florian.rabe@gmail.com}
\thanks{{\lsuper b}The second author was partially supported by a fellowship for Ph.D. research of the German Academic Exchange Service.}

\begin{abstract}
\input{abstract}
\end{abstract}

\keywords{Kripke models, semantics, type theory, dependent types}
\subjclass{F.4.1}
\titlecomment{{\lsuper*}A preliminary version of this paper appeared as \cite{AR:lamkrip:09}.}

\maketitle{}

\section{Introduction and Related Work}\label{sec:mltt:introduction}
  \input{introduction}

\section{Syntax}\label{sec:mltt:syntax}
  \subsection{Grammar}
     \input{syntax}\label{sec:mltt:grammar}

  \subsection{Type System}\label{sec:mltt:rules}

\input{syntax_rules}

\section{Categorical Preliminaries}\label{sec:mltt:preliminaries}
  \input{preliminaries}

\section{Operations on Indexed Sets}\label{sec:mltt:functors}
  \input{functors}

\section{Semantics}\label{sec:mltt:semantics}
  \input{semantics}

\section{Substitution Lemma}\label{sec:mltt:substitution}
	   \input{substitution}

\section{Soundness}\label{sec:mltt:soundness}
	  \input{soundness}
	
\section{Completeness}\label{sec:mltt:completeness}
	  \input{completeness}

\section{Conclusion and Future Work}\label{sec:mltt:conclusion}
  \input{conclusion}

\bibliographystyle{alpha}

\input{LamKrip_long.bbl}
\vspace{-18 pt}
\end{document}

%% file: abstract.tex
It is well-known that simple type theory is complete with respect to non-standard set-valued models. Completeness for standard models only holds with respect to certain extended classes of models, e.g., the class of cartesian closed categories.
Similarly, dependent type theory is complete for locally cartesian closed categories. However, it is usually difficult to establish the coherence of interpretations of dependent type theory, i.e., to show that the interpretations of equal expressions are indeed equal. Several classes of models have been used to remedy this problem.

We contribute to this investigation by giving a semantics that is standard, coherent, and sufficiently general for completeness while remaining relatively easy to compute with. Our models interpret types of Martin-L\"of's extensional dependent type theory as sets indexed over posets or, equivalently, as fibrations over posets. This semantics can be seen as a generalization to dependent type theory of the interpretation of intuitionistic first-order logic in Kripke models. This yields a simple coherent model theory, with respect to which simple and dependent type theory are sound and complete.

%% file: introduction.tex
Martin-L\"of's extensional type theory (\cite{martinlofextensional}, MLTT), is a dependent type theory. Its main characteristic is that there are type-valued function symbols that take terms as input and return types as output. This is enriched with further type constructors such as dependent sum and product. The syntax of dependent type theory is significantly more complex than that of simple type theory because well-formed types and terms and their equalities must be defined in a single joint induction.

The semantics of MLTT is similarly complicated. In \cite{lcccseely}, the connection between MLTT and locally cartesian closed (LCC) categories was first established. LCC categories interpret contexts $\Gamma$ as objects $\sem{\Gamma}{}$, types in context $\Gamma$ as objects in the slice category over $\sem{\Gamma}{}$, substitution as pullback, and dependent sum and product as left and right adjoint to pullback. But there is a difficulty, namely that these three operations are not independent: Substitution of terms into types is associative and commutes with sum and product formation, which is not necessarily the case for the choices of pullbacks and their adjoints. This is known as the coherence or strictness problem and has been studied extensively.
In incoherent models such as in \cite{lccccurien}, equal types are interpreted as isomorphic but not necessarily equal objects.
In \cite{lccccartmell}, coherent models for MLTT are given using categories with attributes.
And in \cite{lccchofmann}, a category with attributes is constructed for every LCC category. Several other model classes and their coherence properties have been studied in, e.g., \cite{lcccstreicher} and \cite{lcccjacobs,lcccjacobs2}. In \cite{pitts00catlog}, an overview is given.

These model classes all have in common that they are rather abstract and have a more complicated structure than general LCC categories. It is clearly desirable to have simpler, more concrete models. But it is a hard problem to equip a given LCC category with choices for pullbacks and adjoints that are both natural and coherent. Our motivation is to find a simple concrete class of LCC categories for which such a choice can be made, and which is still general enough to be complete for MLTT.

Mathematically, our main results can be summarized very simply: Using a theorem from topos theory, it can be shown that MLTT is complete with respect to --- not necessarily coherent --- models in the LCC categories of the form $\Set^P$ for posets $P$, where $\Set$ is the category of sets and mappings. This is equivalent to using presheaves on posets as models, which are often called Kripke models. They were also studied in \cite{lccchofmann2}. For these rather simple models, a solution to the coherence problem can be given. 
$\Set$ can be equipped with a coherent choice of pullback functors, and hence the categories $\Set^P$ can be as well. Deviating subtly from the well-known constructions, we can also make coherent choices for the required adjoints to pullback. Finally, rather than working in the various slices $\slii{\Set^P}{A}$, we use the isomorphism $\slii{\Set^P}{A}\isom\Set^{\grodf{P}{A}}$, where $\grodf{P}{A}$ is the category of elements: Thus we can formulate the semantics of dependent types uniformly in terms of the simple categories of indexed sets $\Set^Q$ for various posets $Q$.
 
In addition to being easy to work with, this has the virtue of capturing the idea that a dependent type $S$ in context $\Gamma$ is in some sense a type-valued function on $\Gamma$: Our models interpret $\Gamma$ as a poset $\sem{\Gamma}{}$ and $S$ as an indexed set $\semc{\Gamma}{S}{}:\sem{\Gamma}{}\arr\Set$.
We speak of Kripke models because these models are a natural extension of the well-known Kripke models for intuitionistic first-order logic (\cite{kripke65intuitionistic}).
Such models are based on a poset $P$ of worlds, and the universe is given as a $P$-indexed set (possibly equipped with $P$-indexed structure).
This can be seen as the special case of our semantics when there is only one type.

In fact, our results are also interesting in the special case of simple type theory (\cite{churchtypes}). Contrary to Henkin models (\cite{henkintypes,mitchell89lambdamodels}), and the models given in \cite{mitchell91kripke}, which like ours use indexed sets on posets, our models are standard: The interpretation $\semc{\Gamma}{S\arr S'}{}$ of the function type is the exponential of $\semc{\Gamma}{S}{}$ and $\semc{\Gamma}{S'}{}$. And contrary to the models in \cite{friedman75equality,simpson95lambdamodels}, our completeness result holds for theories with more than only base types and terms.

A different notion of Kripke-models for dependent type theory is given in \cite{lccclipton}, which is related to \cite{lcccallen}. There, the MLTT types are translated into predicates in an untyped first-order language. The first-order language is then interpreted in a Kripke-model, i.e., there is one indexed universe of which all types are subsets. Such models correspond roughly to non-standard set-theoretical models.

We give the syntax of MLTT in Sect.~\ref{sec:mltt:syntax} and some categorical preliminaries in Sect.~\ref{sec:mltt:preliminaries}. Then we derive the coherent functor choices in Sect.~\ref{sec:mltt:functors} and use them to define the interpretation in Sect.~\ref{sec:mltt:semantics}.
We give our main results regarding the interpretation of substitution, soundness, and completeness in Sect.~\ref{sec:mltt:substitution}, \ref{sec:mltt:soundness}, and~\ref{sec:mltt:completeness}.

%% file: syntax.tex
The basic syntax for MLTT expressions is given by the grammar in Fig.~\ref{fig:mltt:grammar}. The vocabulary of the syntax is declared in signatures and contexts: Signatures $\Sigma$ declare globally accessible names $c$ for constants of type $S$ and names $a$ for type-valued constants with a list $\Gamma$ of argument types. Contexts $\Gamma$ locally declare typed variables $x$.

Substitutions $\gamma$ translate from a context $\Gamma$ to $\Gamma'$ by providing terms in context $\Gamma'$ for the variables in $\Gamma$. Thus, a substitution from $\Gamma$ to $\Gamma'$ can be applied to expressions in context $\Gamma$ and yields expressions in context $\Gamma'$.
Relative to a signature $\Sigma$ and a context $\Gamma$, there are two syntactical classes: types and typed terms.

The base types are the application $\appt{a}{\gamma}$ of a type-valued constant to a list of argument terms $\gamma$ (which we write as a substitution for simplicity).
The composed types are the unit type $\unit$, the identity types $\ident{s}{s'}$, the dependent product types $\S{x}{S}T$, and the dependent function types $\P{x}{S}T$.
Terms are constants $c$, variables $x$, the element $\unitE$ of the unit type, the element $\refl{s}$ of the type $\ident{s}{s}$, pairs $\pair{s}{s'}$, projections $\pi_1(s)$ and $\pi_2(s)$, $\lambda$-abstractions $\lam{x}{S}s$, and function applications $s\;s'$.
We do not need equality axioms $s\meq s'$ because they can be given as constants of type $\ident{s}{s'}$. For simplicity, we omit equality axioms for types.

\begin{figure}
\begin{center}
\begin{tabular}{@{}ll@{\tb::=\tb }l@{}}
Signatures & $\Sigma$ & $\cdot \| \Sigma,\decl{c}{S} \| \Sigma,\declk{a}{\Gamma}$ \\
Contexts & $\Gamma$ & $\cdot \| \Gamma,\decl{x}{S}$ \\
Substitutions & $\gamma$ & $\cdot \| \gamma,\sb{x}{s}$ \\
Types & $S$ & $\appt{a}{\gamma} \| \unit \| \ident{s}{s'} \| \S{x}{S}S' \| \P{x}{S}S'$ \\
Terms & $s$ & $c \| x  \| \unitE \| \refl{s} \| \pair{s}{s'}\| \pi_1(s) \| \pi_2(s) \| \lam{x}{S}s \| s\;s'$ \\
\end{tabular}
\end{center}
\caption{Basic Grammar}\label{fig:mltt:grammar}
\end{figure}

Our formulation of MLTT only uses types and terms. This is different from variants of dependent type theory with kinded type families as in \cite{lambdacube} and \cite{lf}. In particular, in our formulation, the constants $a$ are the only type families, and $a$ itself is not a well-formed expression. All our results extend to the case with kinded type families (see \cite{rabe:thesis:08}).

\begin{definition}[Substitution Application]\label{def:mltt:substitutionapply}
The \defemph{application} of a substitution $\gamma$ to a term, type, or substitution is defined as follows where $\gamma^x$ abbreviates $\gamma,\sb{x}{x}$.

Substitution in terms:
\begin{myeqnarray}[:=]
\subap{\gamma}{c} & c \\
\subap{\gamma}{x} & s & \mfor \sb{x}{s}\minn \gamma\\
\subap{\gamma}{\unitE} & \unitE \\
\subap{\gamma}{\refl{s}} & \refl{\subap{\gamma}{s}} \\
\subap{\gamma}{\pair{s}{s'}} & \pair{\subap{\gamma}{s}}{\subap{\gamma}{s'}} \\
\subap{\gamma}{\pi_1(s)} & \pi_1(\subap{\gamma}{s}) \\
\subap{\gamma}{\pi_2(s)} & \pi_2(\subap{\gamma}{s}) \\
\subap{\gamma}{\lam{x}{S}t} & \lam{x}{\subap{\gamma}{S}}\subapx{\gamma}{x}{t} \\
\subap{\gamma}{f\;s} & \subap{\gamma}{f}\;\subap{\gamma}{s} \\
\end{myeqnarray}

Substitution in types:
\begin{myeqnarray}[:=]
\subap{\gamma}{\unit} & \unit \\
\subap{\gamma}{\ident{s}{s'}} & \ident{\subap{\gamma}{s}}{\subap{\gamma}{s'}} \\
\subap{\gamma}{\S{x}{S}T} & \S{x}{\subap{\gamma}{S}}\subapx{\gamma}{x}{T} \\
\subap{\gamma}{\P{x}{S}T} & \P{x}{\subap{\gamma}{S}}\subapx{\gamma}{x}{T} \\
\subap{\gamma}{\appt{a}{\gammak}} & \appt{a}{\subap{\gamma}{\gammak}} 
\end{myeqnarray}


Substitution in substitutions:
\begin{myeqnarray}[:=]
\subap{\gamma}{\cdot} & \cdot \\
\subap{\gamma}{\sb{x_1}{s_1},\ldots,\sb{x_n}{s_n}} & \sb{x_1}{\subap{\gamma}{s_1}},\ldots,\sb{x_n}{\subap{\gamma}{s_n}} \\
\end{myeqnarray}
Substitution in substitutions is the same as composition of substitutions, and we write $\ö{\delta}{\gamma}$ instead of $\subap{\gamma}{\delta}$.
\end{definition}

%% file: syntax_rules.tex
The judgments defining well-formed syntax are listed in Fig.~\ref{fig:mltt:judgments}.
The typing rules for these judgments are well-known. Our formulation follows roughly \cite{lcccseely}, including the use of extensional identity types. The latter means that the equality judgment for the terms $s$ and $s'$ holds iff the type $\ident{s}{s'}$ is inhabited.

\begin{figure}
\begin{center}
\begin{tabular}{|l|l|}
\hline
Judgment & Intuition \\
\hline
$\issig{\Sigma}$ & $\Sigma$ is a well-formed signature \\
$\iscont{\Sigma}{\Gamma}$ & $\Gamma$ is a well-formed context over $\Sigma$ \\
$\issubs{\Sigma}{\Gamma}{\Gamma'}{\gamma}$ & $\gamma$ is a well-formed substitution over $\Sigma$ from $\Gamma$ to $\Gamma'$ \\
$\ofkind{\Sigma}{\Gamma}{S}{\kity}$ & $S$ is a well-formed type over $\Sigma$ and $\Gamma$ \\
 $\isequal{\Sigma}{\Gamma}{S}{S'}$ & types $S$ and $S'$ are equal over $\Sigma$ and $\Gamma$
\\
$\oftype{\Sigma}{\Gamma}{s}{S}$ & term $s$ is well-formed with type $S$ over $\Sigma$ and $\Gamma$ \\
$\isequal{\Sigma}{\Gamma}{s}{s'}$ & terms $s$ and $s'$ are equal over $\Sigma$ and $\Gamma$ \\
\hline
\end{tabular}
\end{center}
\caption{Judgments}\label{fig:mltt:judgments}
\end{figure}

\begin{example}\label{ex:mltt:cat}
The theory $\excat$ of categories is given by declaring type-valued constants $\exob$ and $\exmor$ and term-valued constants $\exid$ and $\excomp$ such that the following judgments hold

\begin{center}
\begin{tabular}{@{}l@{\;\;}c@{\;\;}l@{\;\;}c@{\;\;}lr@{}}
 $\cdot$ & $\vdash_{\excat}$ & $\exob$ & $:$ & $\kity$ & \\
 $x:\exob,\;y:\exob$ & $\vdash_{\excat}$ & $\exmor\;x\;y$ & $:$ & $\kity$ &  \\
 $x:\exob$ & $\vdash_{\excat}$ & $\exid\;x$ & $:$ & $\exmor\;x\;x$ &  \\
 $x:\exob,\;y:\exob,\;z:\exob,$ \\
 \tb $g:\exmor\;y\;z,\;f:\exmor\;x\;y$ & $\vdash_{\excat}$ & $g\circ f$ & $:$ & $\exmor\;x\;z$ &  \\
 $w:\exob,\;x:\exob,\;y:\exob,\;z:\exob,\;$\\
 \tb $f:\exmor\;w\;x,\;g:\exmor\;x\;y,\;h:\exmor\;y\;z$ & $\vdash_{\excat}$ &
   $h\circ (g\circ f)$ & $\equiv$ & $(h\circ g)\circ f$ &  \\
 $x:\exob,\;y:\exob,\;f:\exmor\;x\;y$ & $\vdash_{\excat}$ & $f\circ \exid\;x$ & $\equiv$ & $f$ &  \\
 $x:\exob,\;y:\exob,\;f:\exmor\;x\;y$ & $\vdash_{\excat}$ & $\exid\;y\circ f$ & $\equiv$ & $f$ &  \\
\end{tabular}
\end{center}

\noindent Here we have used two common abbreviations. (i) $\exmor$ is declared as $\declk{\exmor}{x:\exob,\;y:\exob}$, and we abbreviate the type application $\appt{\exmor}{\sb{x}{s},\;\sb{y}{t}}$ as $\exmor\;s\;t$. (ii) $\circ$ is declared as a constant
 \[\circ\;:\;\P{x}{\exob}\P{y}{\exob}\P{z}{\exob}\P{g}{\exmor\;y\;z}\P{f}{\exmor\;x\;y}\exmor\;x\;z\]
 and we abbreviate $\circ\;x\;y\;z\;g\;f$ as $g\circ f$. This is unambiguous because the values of the first three arguments can be inferred from the types of the last two arguments.

The axioms of a category are declared using the Curry-Howard equivalence (\cite{curry,howard}) of MLTT and intuitionistic first-order logic without negation (\cite{lcccseely}). For example, to obtain right-neutrality, we declare a constant
 \[\exneutr \;:\; \P{x}{\exob}\P{y}{\exob}\P{f}{\exmor\;x\;y}\ident{f\circ \exid\;x}{f}\]
Such a constant yields the corresponding equality judgment above using Rule $e_{\ident{-}{-}}$ from Fig.~\ref{fig:mltt:conversions}.
\end{example}
\medskip

The \defemph{rules} for signatures, contexts, and substitutions are given in Fig.~\ref{fig:mltt:signatures}. A \defemph{signature} is a list of declarations of type-valued constants $a$ or term constants $c$. For example, $\declk{a}{\Gamma}$ means that $a$ can be applied to arguments with types given by $\Gamma$ and returns a type. The domain of a signature is defined by $\dom{\cdot}=\es$, $\dom{\Sigma,\declk{a}{\Gamma}}=\dom{\Sigma}\cup\{a\}$, and $\dom{\Sigma,\decl{c}{S}}=\dom{\Sigma}\cup\{c\}$.


\defemph{Contexts} are similar to signatures except that they only declare variables ranging over terms. The domain of a context is defined as for signatures. A \defemph{substitution} from $\Gamma$ to $\Gamma'$ is a list of terms in context $\Gamma'$ such that each term is typed by the corresponding type in $\Gamma$. Note that in a context $\decl{x_1}{S_1},\ldots,\decl{x_n}{S_n}$, the variable $x_i$ may occur in $S_{i+1},\ldots,S_n$.

\begin{fignd}{mltt:signatures}{Signatures, Contexts, Substitutions}
\ianc{}
     {\issig{\cdot}}
     {\Sigma_{\cdot}}
\tb\tb
\icnc{\issig{\Sigma}}
     {\oftype{\Sigma}{\cdot}{S}{\kity}}
     {c\nin\dom{\Sigma}}
     {\issig{\Sigma,\decl{c}{S}}}
     {\Sigma_c}
\\
\icnc{\issig{\Sigma}}
     {\iscont{\Sigma}{\Gamma'}}
     {a\nin\dom{\Sigma}}
     {\issig{\Sigma,\declk{a}{\Gamma'}}}
     {\Sigma_a}
\\
\ianc{\issig{\Sigma}}
     {\iscont{\Sigma}{\cdot}}
     {\Gamma_{\cdot}}
\tb\tb
\icnc{\iscont{\Sigma}{\Gamma}}
     {\oftype{\Sigma}{\Gamma}{S}{\kity}}
     {x\nin\dom{\Gamma}}
     {\iscont{\Sigma}{\Gamma,\decl{x}{S}}}
     {\Gamma_x}
\\
\ianc{\iscont{\Sigma}{\Gamma'}}
     {\issubs{\Sigma}{\cdot}{\Gamma'}{\cdot}}
     {\sigma_{\cdot}}
\tb\tb
\icnc{\issubs{\Sigma}{\Gamma}{\Gamma'}{\gamma}}
     {\ofkind{\Sigma}{\Gamma}{S}{\kity}}
     {\oftype{\Sigma}{\Gamma'}{s}{\subap{\gamma}{S}}}
     {\issubs{\Sigma}{\Gamma,\decl{x}{S}}{\Gamma'}{\gamma,\sb{x}{s}}}
     {\sigma_x}
\end{fignd}



\begin{fignd}{mltt:types}{Types}
\ibnc{\declk{a}{\Gammak}\minn \Sigma}
     {\issubs{\Sigma}{\Gammak}{\Gamma}{\gammak}}
     {\ofkind{\Sigma}{\Gamma}{\appt{a}{\gammak}}{\kity}}
     {T_\op{app}}
\\
\ianc{\iscont{\Sigma}{\Gamma}}
     {\ofkind{\Sigma}{\Gamma}{\unit}{\kity}}
     {T_\unit}
\tb\tb
\ibnc{\oftype{\Sigma}{\Gamma}{s}{S}}
     {\oftype{\Sigma}{\Gamma}{s'}{S}}
     {\ofkind{\Sigma}{\Gamma}{\ident{s}{s'}}{\kity}}
     {T_{\ident{-}{-}}}
\\
\ianc{\ofkind{\Sigma}{\Gamma,\decl{x}{S}}{T}{\kity}}
     {\ofkind{\Sigma}{\Gamma}{\S{x}{S}T}{\kity}}
     {T_{\Sigma}}
\tb\tb
\ianc{\ofkind{\Sigma}{\Gamma,\decl{x}{S}}{T}{\kity}}
     {\ofkind{\Sigma}{\Gamma}{\P{x}{S}T}{\kity}}
     {T_{\Pi}}
\end{fignd}

\begin{fignd}{mltt:terms}{Terms}
\ibnc{\decl{c}{S}\minn \Sigma}
     {\iscont{\Sigma}{\Gamma}}
     {\oftype{\Sigma}{\Gamma}{c}{S}}
     {t_c}
\tb\tb
\ibnc{\iscont{\Sigma}{\Gamma}}
     {\decl{x}{S}\minn \Gamma}
     {\oftype{\Sigma}{\Gamma}{x}{S}}
     {t_x}
\\
\ianc{\iscont{\Sigma}{\Gamma}}
     {\oftype{\Sigma}{\Gamma}{\unitE}{\unit}}
     {t_{\unitE}}
\tb\tb
\ianc{\oftype{\Sigma}{\Gamma}{s}{S}}
     {\oftype{\Sigma}{\Gamma}{\refl{s}}{\ident{s}{s}}}
     {t_{\refl{-}}}
\\
\icnc{\oftype{\Sigma}{\Gamma}{s}{S}}
     {\oftype{\Sigma}{\Gamma,\decl{x}{S}}{T}{\kity}}
     {\oftype{\Sigma}{\Gamma}{t}{\subapo{T}{x}{s}}}
     {\oftype{\Sigma}{\Gamma}{\pair{s}{t}}{\S{x}{S}T}}
     {t_{\pair{-}{-}}}
\\
\ianc{\oftype{\Sigma}{\Gamma}{u}{\S{x}{S}T}}
     {\oftype{\Sigma}{\Gamma}{\pi_1(u)}{S}}
     {t_{\pi_1}}
\tb\tb
\ianc{\oftype{\Sigma}{\Gamma}{u}{\S{x}{S}T}}
     {\oftype{\Sigma}{\Gamma}{\pi_2(u)}{\subapo{T}{x}{\pi_1(s)}}}
     {t_{\pi_2}}
\\
\ianc{\oftype{\Sigma}{\Gamma,\decl{x}{S}}{t}{T}}
     {\oftype{\Sigma}{\Gamma}{\lam{x}{S}t}{\P{x}{S}T}}
     {t_{\lambda}}
\tb\tb
\ibnc{\oftype{\Sigma}{\Gamma}{f}{\P{x}{S}T}}
     {\oftype{\Sigma}{\Gamma}{s}{S}}
     {\oftype{\Sigma}{\Gamma}{f\;s}{\subapo{T}{x}{s}}}
     {t_{\op{app}}}
\end{fignd}

Fig.~\ref{fig:mltt:types} gives the formation rules for \defemph{types}. In context $\Gamma$, an application $\appt{a}{\gammak}$ of a type constructor $\declk{a}{\Gammak}$ to a substitution $\gammak$ from $\Gammak$ into $\Gamma$, means that $\gammak$ provides a list of terms as arguments to $a$.

Fig.~\ref{fig:mltt:terms} gives the \defemph{term} formation rules. For the case where only one variable is to be substituted in an expression $e$ in context $\Gamma,\decl{x}{S}$, we define
 \[\subapo{e}{x}{s}:=\subap{(\id{\Gamma},\sb{x}{s})}{e}.\] We have the following subexpression property: $\oftype{\Sigma}{\Gamma}{s}{S}$ implies $\ofkind{\Sigma}{\Gamma}{S}{\kity}$ implies $\iscont{\Sigma}{\Gamma}$ implies $\issig{\Sigma}$.

Fig.~\ref{fig:mltt:conversions} gives the congruence and conversion rules for the \defemph{equality of terms}. $\eta$-conversion, reflexivity, symmetry, transitivity, and congruence rules for the other term constructors are omitted because they are derivable or admissible. In particular, $\eta$-conversion is implied by functional extensionality $e_{\op{funcext}}$. The rules have extra premises ensuring well-formedness of subexpressions, but these are elided for
ease of reading, i.e., we assume that all terms occurring in Fig.~\ref{fig:mltt:conversions} are well-formed without making that explicit in the rules.

Finally, Fig.~\ref{fig:mltt:equalitytypes} gives a simple axiomatization of the equality of types. Note that equality of types is decidable iff the equality of terms is.

\begin{fignd}{mltt:conversions}{Equality of Terms}
\ianc{\oftype{\Sigma}{\Gamma}{v}{\ident{s}{s'}}}
     {\isequal{\Sigma}{\Gamma}{s}{s'}}
     {e_{\ident{-}{-}}}
\tb\tb
\ibnc{\oftype{\Sigma}{\Gamma}{v}{\ident{s}{s'}}}
     {\oftype{\Sigma}{\Gamma}{v'}{\ident{s}{s'}}}
     {\isequal{\Sigma}{\Gamma}{v}{v'}}
     {e_{\op{id-uniq}}}
\\
\ianc{\oftype{\Sigma}{\Gamma}{s}{\unit}}
     {\isequal{\Sigma}{\Gamma}{s}{\unitE}}
     {e_{\unitE}}
\tb\tb
\ianc{}
     {\isequal{\Sigma}{\Gamma}{\rang{\pi_1(u),\pi_2(u)}}{u}}
     {e_{\pair{-}{-}}}
\\
\ianc{}
     {\isequal{\Sigma}{\Gamma}{\pi_1(\pair{s}{s'})}{s}}
     {e_{\pi_1}}
\tb\tb
\ianc{}
     {\isequal{\Sigma}{\Gamma}{\pi_1(\pair{s}{s'})}{s'}}
     {e_{\pi_2}}
\\
\ianc{}
     {\isequal{\Sigma}{\Gamma}{(\lam{x}{S}t)\;s}{\subapo{t}{x}{s}}}
     {e_\beta}
\tb\tb
\ibnc{\isequal{\Sigma}{\Gamma}{f}{f'}}
     {\isequal{\Sigma}{\Gamma}{s}{s'}}
     {\isequal{\Sigma}{\Gamma}{f\;s}{f'\;s'}}
     {e_{\op{app}}}
\\
\icnc{\oftype{\Sigma}{\Gamma}{f}{\P{x}{S}T}}
     {\oftype{\Sigma}{\Gamma}{f'}{\P{x}{S}T}}
     {\isequal{\Sigma}{\Gamma,\decl{y}{S}}{f\;y}{f'\;y}}
     {\isequal{\Sigma}{\Gamma}{f}{f'}}
     {e_{\op{funcext}}}
\\
\icnc{\oftype{\Sigma}{\Gamma}{s}{S}}
     {\isequal{\Sigma}{\Gamma}{s}{s'}}
     {\isequal{\Sigma}{\Gamma}{S}{S'}}
     {\oftype{\Sigma}{\Gamma}{s'}{S'}}
     {e_{\op{typing}}}
\end{fignd}

\begin{fignd}{mltt:equalitytypes}{Equality of Types}
\ibnc{{\gamma=\sb{x_1}{s_1},\ldots,\sb{x_n}{s_n}
      \atop
      \gamma'=\sb{x_1}{s'_1},\ldots,\sb{x_n}{s'_n}}}
     {\isequal{\Sigma}{\Gamma}{s_i}{s'_i}\;\mfor i=1,\ldots,n}
     {\isequal{\Sigma}{\Gamma}{\appt{a}{\gamma}}{\appt{a}{\gamma'}}}
     {E_a}
\\
\ianc{}
     {\isequal{\Sigma}{\Gamma}{\unit}{\unit}}
     {E_\unit}
\tb\tb
\ibnc{\isequal{\Sigma}{\Gamma}{s_1}{s'_1}}
     {\isequal{\Sigma}{\Gamma}{s_2}{s'_2}}
     {\isequal{\Sigma}{\Gamma}{\ident{s_1}{s_2}}{\ident{s'_1}{s'_2}}}
     {E_{\ident{-}{-}}}
\\
\ibnc{\isequal{\Sigma}{\Gamma}{S}{S'}}
     {\isequal{\Sigma}{\Gamma,\decl{x}{S}}{T}{T'}}
     {\isequal{\Sigma}{\Gamma}{\S{x}{S}T}{\S{x}{S'}T'}}
     {E_\Sigma}
\tb\tb
\ibnc{\isequal{\Sigma}{\Gamma}{S}{S'}}
     {\isequal{\Sigma}{\Gamma,\decl{x}{S}}{T}{T'}}
     {\isequal{\Sigma}{\Gamma}{\P{x}{S}T}{\P{x}{S'}T'}}
     {E_\Pi}
\end{fignd}
\medskip

Parallel to Def.~\ref{def:mltt:substitutionapply}, we obtain the following basic property of substitutions by a straightforward induction on derivations:

\begin{lemma}\label{lem:mltt:substitutionapply}
Assume $\issubs{\Sigma}{\Gamma}{\Gamma'}{\gamma}$. Then:
\begin{center}
\begin{tabular}{@{if\tb}l@{\tb then\tb}l}
	$\issubs{\Sigma}{\Delta}{\Gamma}{\delta}$ & $\issubs{\Sigma}{\Delta}{\Gamma'}{\ö{\delta}{\gamma}}$, \\
	$\ofkind{\Sigma}{\Gamma}{S}{\kity}$       & $\ofkind{\Sigma}{\Gamma'}{\subap{\gamma}{S}}{\kity}$,   \\
	$\oftype{\Sigma}{\Gamma}{s}{S}$           & $\oftype{\Sigma}{\Gamma'}{\subap{\gamma}{s}}{\subap{\gamma}{S}}$.
\end{tabular}
\end{center}
\end{lemma}



%% file: preliminaries.tex
In this section, we repeat some well-known definitions and results about indexed sets and fibrations over posets (see, e.g., \cite{johnstone}). We assume the basic notions of category theory (see, e.g., \cite{categories}). We use a set-theoretical pairing function $(a,b)$ and define tuples as left-associatively nested pairs, i.e., $(a_1,a_2,\ldots,a_n)$ abbreviates $(\ldots(a_1,a_2),\ldots,a_n)$.

\begin{definition}[Indexed Sets]
$\Poset$ denotes the category of partially ordered sets. We treat posets as categories and write $\mr{p}{p'}$ for the uniquely determined morphism $p\arr p'$. If $P$ is a poset, $\Pres{P}$ denotes the category of functors $P\arr\Set$ and natural transformations. These functors are also called $P$-\defemph{indexed sets}.
\end{definition}
We denote the constant $P$-indexed set that maps each $p\in P$ to $\{\emptytup\}$ by $\One_P$.
It is often convenient to replace an indexed set $A$ over $P$ with a poset formed from the disjoint union of all sets $A(p)$ for $p\in P$. 
This is a special case of the category of elements, a construction due to Mac Lane (\cite{sheaves}) that is sometimes also called the Grothendieck construction.

\begin{definition}[Category of Elements]\label{def:mltt:groethendieck}
For an indexed set $A$ over $P$, we define a poset $\grodf{P}{A}:=\{(p,a)\|p\in P, a\in A(p)\}$ with
 \[(p,a)\leq(p',a')\tb\miff\tb p\leq p'\mand A(\mr{p}{p'})(a)=a'.\]
We also write $\grod{A}$ instead of $\grodf{P}{A}$ if $P$ is clear from the context.
\end{definition}

Using the category of elements, we can work with sets indexed by indexed sets: We write $\indset{P,A}$ if $A$ is an indexed set over $P$, and $\indset{P,A,B}$ if additionally $B$ is an indexed set over $\grodf{P}{A}$, etc.

\begin{definition}
Assume $\indset{P,A,B}$. We define an indexed set $\indset{P,(\grot{A}{B})}$ by
 \[(\grot{A}{B})(p)=\{(a,b)\|a\in A(p), b\in B(p,a)\}\]
 and
 \[(\grot{A}{B})(\mr{p}{p'}):(a,b)\mapsto\big(a',B\big(\mr{(p,a)}{(p',a')}\big)(b)\big)\tb\mfor a'=A(\mr{p}{p'})(a).\]
And we define a natural transformation $\grop{B}:\grot{A}{B}\arr A$ by
 \[(\grop{B})_p:(a,b)\mapsto a.\]
\end{definition}

The following definition introduces discrete opfibrations; for brevity, we will refer to them as ``fibrations'' in the sequel. Using the axiom of choice, these are necessarily split.
\begin{definition}[Fibrations]
A \defemph{fibration} over a poset $P$ is a functor $f:Q\arr P$ for a poset $Q$ with the following property: For all $p'\in P$ and $q\in Q$ such that $f(q)\leq p'$, there is a unique $q'\in Q$ such that $q\leq q'$ and $f(q')=p'$.
We call $f$ \defemph{canonical} iff $f$ is the first projection of $Q=\grodf{P}{A}$ for some $\indset{P,A}$.
\end{definition}

For every indexed set $A$ over $P$, the first projection $\grodf{P}{A}\arr P$ is a (canonical) fibration. Conversely, every fibration $f:Q\arr P$ defines an indexed set over $P$ by mapping $p\in P$ to its preimage $f^{-1}(p)\sq Q$ and $\mr{p}{p'}$ to the obvious function. This leads to a well-known equivalence of indexed sets and fibrations over $P$. If we only consider canonical fibrations, we obtain an isomorphism as follows.

\begin{lemma}\label{lem:mltt:equiv}
If we restrict the objects of $\slii{\Poset}{P}$ to be canonical fibrations and the morphisms to be (arbitrary) fibrations, we obtain the full subcategory $\Fib{P}$ of $\slii{\Poset}{P}$. There are isomorphisms
\[\groF{-}:\Pres{P}\arr\Fib{P} \tb\mand\tb \groI{-}:\Fib{P}\arr \Pres{P}.\]
\end{lemma}
\begin{proof}
It is straightforward to show that $\Fib{P}$ is a full subcategory:
The identity in $\Poset$ and the composition of two fibrations are fibrations. Thus, it only remains to show that if $\ö{\phi}{f}=f'$ in $\Poset$ where $f$ and $f'$ are fibrations and $\phi$ is a morphism in $\Poset$, then $\phi$ is a fibration as well. This is easy.

For $A:P\arr\Set$, we define the fibration $\groF{A}:\grodf{P}{A}\arr P$ by $(p,a)\mapsto p$. And for a natural transformation $\eta:A\arr A'$, we define the fibration $\groF{\eta}:\grodf{P}{A}\arr\grodf{P}{A'}$ satisfying $\ö{\groF{\eta}}{\groF{A}}=\groF{A'}$ by $(p,a)\mapsto(p,\eta_p(a))$.

For $f:Q\arr P$, we obtain an indexed set using the fact that $f$ is canonical. More concretely, we define $\groI{f}(p):=\{a\|f(p,a)=p\}$ and $\groI{f}(\mr{p}{p'}):a\mapsto a'$ where $a'$ is the uniquely determined element such that $(p,a)\leq (p',a')\in Q$. And for a morphism $\phi$ between fibrations $f:Q\arr P$ and $f':Q'\arr P$, we define a natural transformation $\groI{\phi}:\groI{f}\arr \groI{f'}$ by $\groI{\phi}_p:a\mapsto a'$ where $a'$ is such that $\phi(p,a)=(p,a')$.

Then it is easy to compute that $I$ and $F$ are mutually inverse functors.
\myqed\end{proof}

\begin{definition}[Indexed Elements]\label{def:mltt:indelements}
Assume $\indset{P,A}$. The $P$-\defemph{indexed elements} of $A$ are given by
\[\STerms{A}:=\big\{\big(a_p\in A(p)\big)_{p\in P}\| a_{p'}=A(\mr{p}{p'})(a_p)\mbox{ whenever }p\leq p'\big\}.\]
\end{definition}
Then the indexed elements of $A$ are in bijection with the natural transformations $\One_P\arr A$.
For $a\in\STerms{A}$, we will write $\groF{a}$ for the fibration $P\arr\grod{A}$ mapping $p$ to $(p,a_p)$.
$\groF{a}$ is a section of $\groF{A}$, and indexed elements are also called global sections.
 
\begin{example}
We exemplify the introduced notions by Fig.~\ref{fig:mltt:indexedsets}. $P$ is a totally ordered set visualized as a horizontal line with two elements $p_1\leq p_2\in P$. For $\indset{P,A}$, $\grod{A}$ becomes a blob over $P$. The sets $A(p_i)$ correspond to the vertical lines in $\grod{A}$, and $a_i\in A(p_i)$. The action of $A(\mr{p}{p'})$ and the poset structure of $\grod{A}$ are horizontal: If we assume $A(\mr{p_1}{p_2}):a_1\mapsto a_2$, then $(p_1,a_1)\leq (p_2,a_2)$ in $\grod{A}$. Finally, the action of $\groF{A}$ is vertical: $\groF{A}$ maps $(p_i,a_i)$ to $p_i$. Note that our intuitive visualization is not meant to indicate that the sets $A(p_i)$ must be in bijection or that the mapping $A(\mr{p_1}{p_2})$ must be injective or surjective. 

Similarly, for $\indset{P,A,B}$, $\grod{B}$ becomes a three-dimensional blob over $\grod{A}$. The sets $B(p_i,a_i)$ correspond to the dotted lines. Again the action of $B(\mr{(p_1,a_1)}{(p_2,a_2)})$ and the poset structure of $\grod{B}$ are horizontal:
\[b_i\in B(p_i,a_i) \tb\mand\tb B(\mr{(p_1,a_1)}{(p_2,a_2)}):b_1\mapsto b_2\]
and $\groF{B}$ projects vertically from $\grod{B}$ to $\grod{A}$.

Similarly, we have
\[(a_i,b_i)\in (\grot{A}{B})(p_i) \tb\mand\tb (\grot{A}{B})(\mr{p_1}{p_2}):(a_1,b_1)\mapsto (a_2,b_2)\]
Thus, the sets $(\grot{A}{B})(p_i)$ correspond to the two-dimensional gray areas. 
The sets 
$\grodf{P}{(\grot{A}{B})}$ and $\grodf{\grodf{P}{A}}{B}$ are isomorphic, and their elements differ only in the bracketing:
\[(p_i,(a_i,b_i))\in\grodf{P}{(\grot{A}{B})} \tb\mand\tb((p_i,a_i),b_i)\in\grodf{\grodf{P}{A}}{B}.\]
Up to this isomorphism, the projection $\groF{\grot{A}{B}}$ is the composite $\ö{\groF{B}}{\groF{A}}$.

Indexed elements $a\in\STerms{A}$ are families $(a_p)_{p\in P}$ and correspond to horizontal curves through $\grod{A}$ such that $\groF{a}$ is a section of $\groF{A}$.
Indexed elements of $B$ correspond to two-dimensional vertical areas in $\grod{B}$ (intersecting each line parallel to the dotted lines exactly once), and indexed elements of $\grot{A}{B}$ correspond to horizontal curves in $\grod{B}$ (intersecting each area parallel to the gray areas exactly once).

Finally the condition that indexed elements are natural transformations can be visualized as follows: The indexed elements $a\in\STerms{A}$ are exactly those horizontal curves that arise if a line is drawn from $(p,a)$ to $(p',a')$ whenever $(p,a)\leq (p',a')$. There may be multiple such curves going through a point $(p,a)$, but they must coincide to the right of $(p,a)$. Moreover, $(p,a)\leq (p',a')$ holds iff $(p,a)$ is to the left of $(p',a')$ on the same curve. In particular, if $P$ has a least element $p_0$, we obtain exactly one such curve for every element of $A(p_0)$.
\end{example}

\begin{figure}[tbh]
\begin{center}
\begin{tikzpicture}[scale=0.7]
\filldraw[gray] (1,6) -- (1,8) -- (3,11) -- (3,9) -- cycle;
\filldraw[lightgray] (3,6) -- (3,8) -- (5,11) -- (5,9) -- cycle;
\draw (4,6) rectangle (0,8) node[left] {$\grod{(\grot{A}{B})}\isom\grod{B}$};
\draw[dashed] (4,6) -- (6,9) -- (6,11) -- (4,8);
\draw[dashed] (0,8) -- (2,11) -- (6,11);
\draw[dotted] (1,7.5) -- plot[mark=+,mark options=black] coordinates {(2,9)} node[left,color=black] {$(p_1,a_1,b_1)$} -- (3,10.5);
\draw[dotted] (3,7.5) -- plot[mark=+,mark options=black] coordinates {(4,9)} node[right,color=black] {$(p_2,a_2,b_2)$} -- (5,10.5);

\draw[-\arrowtip] (2,5.5) -- node[right] {$\groF{B}$} (2,4.5);

\draw (4,2) rectangle (0,4) node[left] {$\grod{A}$}; 
\draw[gray] (1,2) -- plot[mark=x,mark options=black] coordinates {(1,3.5)} node[right,color=black] {$(p_1,a_1)$} -- (1,4);
\draw[lightgray] (3,2) -- plot[mark=x,mark options=black] coordinates {(3,3.5)} node[right,color=black] {$(p_2,a_2)$} -- (3,4);

\draw[-\arrowtip] (2,1.5) -- node[right] {$\groF{A}$} (2,0.5);

\draw (4,0) -- plot[mark=x] coordinates {(1,0)} node[below] {$p_1$} -- plot[mark=x] coordinates {(3,0)} node[below] {$p_2$} -- (0,0) node[left] {$P$};

\node at (9,6.5)
    {$b_i\in B(p_i,a_i),\;B(\mr{(p_1,a_1)}{(p_2,a_2)})=b_2$};
\node at (9,6)
    {$(p_1,a_1,b_1)\leq (p_2,a_2,b_2)$};
\node at (9,5.5) {$(a_i,b_i)\in(\grot{A}{B})(p_i)$};
\node at (9,3.5)
    {$a_i\in A(p_i),\;A(\mr{p_1}{p_2})(a_1)=a_2$};
\node at (9,3)
    {$(p_1,a_1)\leq (p_2,a_2)$};
\node at (9,0)
    {$p_1\leq p_2$};
 \end{tikzpicture}
\end{center}
\caption{Indexed Sets and Fibrations}\label{fig:mltt:indexedsets}
\end{figure}

\begin{example}\label{ex:mltt:mltt}
Let $\Sign$ be the set of well-formed signatures of MLTT (or of any other type theory for that matter). $\Sign$ is a poset under inclusion $\sq$ of signatures. Let $\Con(\Sigma)$ be the set of well-formed contexts over $\Sigma$, and let $\Con(\Sigma\sq \Sigma'):\Con(\Sigma)\harr\Con(\Sigma')$ be an inclusion. Then $\indset{\Sign,\Con}$, and the tuple assigning the empty context to every signature is an example of an indexed element of $\Con$.

$\grodf{\Sign}{\Con}$ is the set of pairs $(\Sigma,\Gamma)$ such that $\iscont{\Sigma}{\Gamma}$, and $(\Sigma,\Gamma)\leq(\Sigma',\Gamma')$ iff $\Sigma\sq\Sigma'$ and $\Gamma=\Gamma'$. Let $\Tp(\Sigma,\Gamma)$ be the set of types $S$ such that $\ofkind{\Sigma}{\Gamma}{S}{\kity}$. $\Tp$ becomes an indexed set $\indset{\Sign,\Con,\Tp}$ by defining $\Tp((\Sigma,\Gamma)\leq(\Sigma',\Gamma))$ to be an inclusion. The tuple assigning $\unit$ to every pair $(\Sigma,\Gamma)$ is an example of an indexed element of $\Tp$.
\end{example}

We will use Lem.~\ref{lem:mltt:equiv} frequently to switch between indexed sets and fibrations, as convenient. In particular, we will use the following two corollaries.
\begin{lemma}\label{lem:mltt:slices}\label{lem:mltt:indexedelements}
Assume $\indset{P,A}$. Then
\[\STerms{A}\isom\Hom_{\Fib{P}}(\id{P},\groF{A})=\{f:P\arr\grodf{P}{A}\|\ö{f}{\groF{A}}=\id{P}\}.\]
and
\[\slii{\Set^P}{A}\isom \Set^\grod{A}\]
\end{lemma}
\begin{proof}
Both claims follow from Lem.~\ref{lem:mltt:equiv} by using $\STerms{A}\isom\Hom_{\Pres{P}}(\One_P,A)$ as well as $\slii{\Fib{P}}{\groF{A}}\cong\Fib{\grodf{P}{A}}$, respectively.
\myqed\end{proof}

Finally, as usual, we say that a category is \defemph{locally cartesian closed} (LCC) if it and all of its slice categories are cartesian closed (in particular, it has a terminal object). Then we have the following well-known result.
\begin{lemma}\label{lem:mltt:lccc}
$\Pres{P}$ is LCC.
\end{lemma}
\input{proof_lccc}

%% file: proof_lccc.tex
\begin{proof}
The terminal object is given by $\One_P$. The product is taken pointwise: $A\times B:p\marr A(p)\times B(p)$ and similarly for morphisms. The exponential object is given by: $B^A: p\marr \Hom_{\Pres{P^p}}(A^p,B^p)$ where $A^p$ and $B^p$ are as $A$ and $B$ but restricted to $P^p:=\{p'\in P\|p\leq p'\}$. $B^A(\mr{p}{p'})$ maps a natural transformation, which is a family of mappings over $P^p$, to its restriction to $P^{p'}$. This proves that $\Set^P$ and so also $\Fib{P}$ is cartesian closed for any $P$. By Lem.~\ref{lem:mltt:slices}, we obtain the same for all slice categories.
\myqed\end{proof}

%% file: functors.tex
Because $\Pres{P}$ is LCC, we know that it has pullbacks and that the pullback along a fixed natural transformation has left and right adjoints (see, e.g., \cite{johnstone}). However, these functors are only unique up to isomorphism, and it is non-trivial to pick coherent choices for them.

\paragraph{Pullbacks}
Assume $\indset{P,A_1}$ and $\indset{P,A_2}$ and a natural transformation $h:A_2\arr A_1$. The pullback along $h$ is a functor $\slii{\Pres{P}}{A_1}\arr\slii{\Pres{P}}{A_2}$. Using Lem.~\ref{lem:mltt:slices}, we can avoid dealing with slice categories of $\Pres{P}$ and instead give a functor
\[\pb{h}{}: \Pres{\grod{A_1}}\arr\Pres{\grod{A_2}},\]
which we also call the pullback along $h$. The functor $\pb{h}{}$ is given by precomposition:
\begin{definition}\label{def:mltt:pullback}
Assume $A_1$ and $A_2$ indexed over $P$, and a natural transformation $h:A_2\arr A_1$. Then for $B\in\Pres{\grod{A_1}}$, we put 
 \[\pb{h}{B}:=\ö{\groF{h}}{B}\in\Pres{\grod{A_2}},\]
where, as in Lem.~\ref{lem:mltt:equiv}, $\groF{h}:\grodf{P}{A_2}\arr\grodf{P}{A_1}$.
The action of $\pb{h}{}$ on morphisms is defined similarly by composing a natural transformation $\beta:B\arr B'$ with the functor $\groF{h}$: $\pb{h}{\beta}:=\ö{\groF{h}}{\beta}$.
Finally, we define a natural transformation between $P$-indexed sets by
 \[\pbf{h}{B}:\grot{A_2}{\pb{h}{B}}\arr \grot{A_1}{B},\tb(\pbf{h}{B})_p:(a_2,b)\mapsto(h_p(a_2),b).\]
\end{definition}\medskip

\noindent The application of $\pbf{h}{B}$ is independent of $B$, which
 is only needed in the notation to determine the domain and codomain
 of $\pbf{h}{B}$.\newpage

\begin{lemma}[Pullbacks]\label{lem:mltt:pullback}
In the situation of Def.~\ref{def:mltt:pullback}, the following is a pullback in $\Pres{P}$.
\begin{center}
\begin{tikzpicture}
\node (B1) at (0,2)
    {$\grot{A_2}{\pb{h}{B}}$};
\node (B2) at (3,2)
    {$\grot{A_1}{B}$};
\node (C1) at (0,0)
    {$A_2$};
\node (C2) at (3,0)
    {$A_1$};
\draw[-\arrowtip](B1) -- node[above] {$\pbf{h}{B}$} (B2);
\draw[-\arrowtip](C1) -- node[above] {$h$} (C2);
\draw[-\arrowtip](B1) -- node[right] {$\grop{\pb{h}{B}}$} (C1);
\draw[-\arrowtip](B2) -- node[left] {$\grop{B}$} (C2);
\end{tikzpicture}
\end{center}
Furthermore, we have the following coherence properties for every natural transformation $g:A_3\arr A_2$:
\[\pb{(\id{A_1})}{B}=B,\tb \pbf{\id{A_1}}{B}=\id{\grot{A_1}{B}},\]
\[\pb{(\ö{g}{h})}{B}=\pb{g}{(\pb{h}{B})},\tb \pbf{(\ö{g}{h})}{B}=\ö{(\pbf{g}{\pb{h}{B}})}{(\pbf{h}{B})}.\]
\end{lemma}
\begin{proof}
The following is a pullback in $\Poset$:
\begin{center}
\begin{tikzpicture}
\node (B1) at (0,2)
    {$\grod{\grot{A_2}{\pb{h}{B}}}$};
\node (B2) at (3,2)
    {$\grod{\grot{A_1}{B}}$};
\node (C1) at (0,0)
    {$\grod{A_2}$};
\node (C2) at (3,0)
    {$\grod{A_1}$};
\node (b1) at (5,2)
    {$(p,(a_2,b))$};
\node (b2) at (8.5,2)
    {$(p,(h_p(a_2),b))$};
\node (c1) at (5,0)
    {$(p,a_2)$};
\node (c2) at (8.5,0)
    {$(p,h_p(a_2))$};
\draw[-\arrowtip](B1) -- node[above] {$\groF{\pbf{h}{B}}$} (B2);
\draw[-\arrowtip](C1) -- node[above] {$\groF{h}$} (C2);
\draw[-\arrowtip](B1) -- node[right] {$\groF{\grop{\pb{h}{B}}}$} (C1);
\draw[-\arrowtip](B2) -- node[left] {$\groF{\grop{B}}$} (C2);
\draw[|-\arrowtip](b1) -- node[above] {$\groF{\pbf{h}{B}}$} (b2);
\draw[|-\arrowtip](c1) -- node[above] {$\groF{h}$} (c2);
\draw[|-\arrowtip](b1) -- node[right] {$\groF{\grop{\pb{h}{B}}}$} (c1);
\draw[|-\arrowtip](b2) -- node[left] {$\groF{\grop{B}}$} (c2);
\end{tikzpicture}
\end{center}
If we turn this square into a cocone on $P$ by adding the canonical projections $\groF{A_2}$ and $\groF{A_1}$, it becomes a pullback in $\Fib{P}$. Then the result follows by Lem.~\ref{lem:mltt:equiv}. The coherence properties can be verified by simple computations.
\myqed\end{proof}

Equivalently, using the terminology of \cite{pitts00catlog}, we can say that for every $P$ the tuple \[(\Pres{P},\Pres{\grod{A}},\grot{A}{B},\grop{B},\pb{h}{B},\pbf{h}{B})\] forms a type category (where $A$, $B$, $h$ indicate arbitrary arguments). Then giving coherent adjoints to the pullback functor shows that this type category admits dependent sums and products.

\paragraph{Adjoints}
To interpret MLTT, the adjoints to $\pb{h}{}$, where $h:A_2\arr A_1$, are only needed if $h$ is a projection, i.e., $A_1:=A$, $A_2:=\grot{A}{B}$, and $h:=\grop{B}$ for some $\indset{P,A,B}$. We only give adjoint functors for this special case because we use this restriction when defining the right adjoint. Thus, we give functors
\[\pgro{B}{},\spi{B}{}: \Pres{\grod{\grot{A}{B}}}\arr\Pres{\grod{A}}
 \;\msuchthat\; \pgro{B}{}\dashv\pb{\grop{B}}{}\dashv\spi{B}{}\]
in Def.~\ref{def:mltt:ladj} and~\ref{def:mltt:radj}, respectively.
These functors will satisfy the coherence properties
\[\pb{g}{(\pgro{B}{C})}=\pgro{\pb{g}{B}}{\pb{(\pbf{g}{B})}{C}} \;\;\mand\;\;
  \pb{g}{(\spi{B}{C})}=\spi{\pb{g}{B}}{\pb{(\pbf{g}{B})}{C}}
\]
for every $g:A'\arr A$, which we prove in Lem.~\ref{lem:mltt:ladj} and~\ref{lem:mltt:radj}, respectively.

\begin{definition}\label{def:mltt:ladj}
We define the functor $\pgro{B}{}$ as follows.
For an object $C$, we put $\pgro{B}{C}:=\grot{B}{(\ö{\assoc}{C})}$ where $\assoc$ maps elements $((p,a),b)\in\grod{B}$ to $(p,(a,b))\in\grod{\grot{A}{B}}$; and for a morphism, i.e., a natural transformation $\eta:C\arr C'$, we put
\[(\pgro{B}{\eta})_{(p,a)}:(b,c)\mapsto(b,\eta_{(p,(a,b))}(c))\tb\mfor (p,a)\in\grod{A}.\]
\end{definition}

\begin{lemma}[Left Adjoint]\label{lem:mltt:ladj}
$\pgro{B}{}$ is left adjoint to $\pb{\grop{B}}{}$. Furthermore, for any natural transformation $g:A'\arr A$, we have the following coherence property (the Beck-Chevalley condition)
\[\pb{g}{(\pgro{B}{C})}=\pgro{\pb{g}{B}}{\pb{(\pbf{g}{B})}{C}}.\]
\end{lemma}
\begin{proof}
It is easy to show that $\pgro{B}{}$ is isomorphic to composition along $\grop{B}$, for which the adjointness is well-known. In particular, we have the following diagram in $\Pres{P}$:
\begin{center}
\begin{tikzpicture}[scale=0.95]
\node (C2') at (5,6)
    {$\grot{(\grot{A}{B})}{C}$};
\node (C2) at (8.5,6)
    {$\grot{A}{\pgro{B}{C}}$};
\node (B2) at (5,4)
    {$\grot{A}{B}$};
\node (A2) at (5,2)
    {$A$};
\draw[\arrowtip-\arrowtip](C2) -- node[above] {$\isom$} (C2');
\draw[-\arrowtip](C2) -- node[right] {$\grop{\pgro{B}{C}}$} (A2);
\draw[-\arrowtip](C2') -- node[left] {${\grop{C}}$} (B2);
\draw[-\arrowtip](B2) -- node[left] {${\grop{B}}$} (A2);
\end{tikzpicture}
\end{center}
The coherence can be verified by direct computation.
\myqed\end{proof}

The right adjoint is more complicated. Intuitively, $\spi{B}{C}$ must represent the dependent functions from $B$ to $C$. The naive candidate for this is $\STerms{C}\isom\Hom(\One_{\grod{B}},C)$ (i.e., $\Hom(B,C)$ in the simply-typed case), but this is not a $\grod{A}$-indexed set. There is a well-known construction to remedy this, but we use a subtle modification to achieve coherence, i.e., the corresponding Beck-Chevalley condition. To do that, we need an auxiliary definition.
\begin{definition}\label{def:mltt:radj_aux}
Assume $\indset{P,A,B}$, $\indset{P,(\grot{A}{B}),C}$, and an element $x:=(p,a)\in \grod{A}$. Let $\Ax\in\Pres{P}$ and a natural transformation $\ix:\Ax\arr A$ be given by
 \[\Ax(p')=\cas{\{\emptytup\}\mifc p\leq p'\\\es\mothw}\tb \ix_{p'}:\emptytup\mapsto A(\mr{p}{p'})(a).\]
Then we define indexed sets $\indset{P,\Ax,\Bx}$ and $\indset{P,(\grot{\Ax}{\Bx}),\Cx}$ by:
\[\Bx:=\pb{\ix}{B}, \tb \Cx:=\pb{(\pbf{\ix}{B})}{C}\]
and put $\dx:=\grod{\grot{\Ax}{\Bx}}$ for the domain of $\Cx$.
\end{definition}
Note that $\Ax$ is the Yoneda embedding of $p$ in $\Pres{P}$. The left diagram in Fig.~\ref{fig:mltt:iswithout} shows the involved $P$-indexed sets, the right one gives the actions of the natural transformations for an element $p'\in P$ with $p\leq p'$. Below it will be crucial for coherence that $\Bx$ and $\Cx$ contain tuples in which $a'$ is replaced with $\emptytup$.

\begin{figure}[ht]
\begin{center}
\begin{tikzpicture}[scale=.85]
\node (Cx) at (0,6)
    {$\grotl{\Ax}{\Bx}{\Cx}$};
\node (C) at (4,6)
    {$\grotl{A}{B}{C}$};
\node (Bx) at (0,4)
    {$\grot{\Ax}{\Bx}$};
\node (B) at (4,4)
    {$\grot{A}{B}$};
\node (Pp) at (0,2)
    {$\Ax$};
\node (A) at (4,2)
    {$A$};
\draw[-\arrowtip](Cx) -- node[above=0.2cm] {$\pbf{(\pbf{\ix}{B})}{C}$} (C);
\draw[-\arrowtip](Bx) -- node[above] {$\pbf{\ix}{B}$} (B);
\draw[-\arrowtip](Pp) -- node[above] {$\ix$} (A);
\draw[-\arrowtip](Cx) -- node[right] {$\grop{\Cx}$} (Bx);
\draw[-\arrowtip](Bx) -- node[right] {$\grop{\Bx}$} (Pp);
\draw[-\arrowtip](C) -- node[left] {$\grop{C}$} (B);
\draw[-\arrowtip](B) -- node[left] {$\grop{B}$} (A);
\end{tikzpicture}
\begin{tikzpicture}[scale=.85]
\node (Cx) at (0,6)
    {$(\emptytup,b',c')$};
\node (C) at (2.5,6)
    {$(a',b',c')$};
\node (Bx) at (0,4)
    {$(\emptytup,b')$};
\node (B) at (2.5,4)
    {$(a',b')$};
\node (Pp) at (0,2)
    {$\emptytup$};
\node (A) at (2.5,2)
    {$a'$};
\node at (4,3.5)
    {$x:=(p,a)$};
\node at (4.3,3)
    {$a':=A(\mr{p}{p'})(a)$};
\draw[|-\arrowtip](Cx) -- (C);
\draw[|-\arrowtip](Bx) -- (B);
\draw[|-\arrowtip](Pp) -- (A);
\draw[|-\arrowtip](Cx) -- (Bx);
\draw[|-\arrowtip](Bx) -- (Pp);
\draw[|-\arrowtip](C) -- (B);
\draw[|-\arrowtip](B) -- (A);
\end{tikzpicture}
\caption{The Situation of Def.~\ref{def:mltt:radj_aux}}\label{fig:mltt:iswithout}
\end{center}
\end{figure}

\begin{definition}\label{def:mltt:radj}
Assume $\indset{P,A,B}$. Then we define the functor $\spi{B}{}:\Pres{\grod{\grot{A}{B}}}\arr\Pres{\grod{A}}$ as follows. Firstly, for an object $C$, we put for $x\in\grod{A}$
\[(\spi{B}{C})(x):=\STerms{\Cx}.\]
In particular, $f\in(\spi{B}{C})(x)$ is a family $(f_y)_{y\in \dx}$ with $f_y\in\Cx(y)$. For $x\leq x'\in\grod{A}$, we have $\dx\supseteq \iswithout{d}{x'}$ and put
\[(\spi{B}{C})(\mr{x}{x'}):(f_y)_{y\in \dx}\mapsto (f_{y})_{y\in d^{x'}}.\]
Secondly, for a morphism, i.e., a natural transformation $\eta:C\arr C'$, we define
$\spi{B}{\eta}:\spi{B}{C}\arr\spi{B}{C'}$ as follows: For $x:=(p,a)\in\grod{A}$ and $f\in(\spi{B}{C})(x)$, we define $f':=(\spi{B}{\eta})_{x}(f)\in(\spi{B}{C'})(x)$ by
\[f'_{(p',(\emptytup,b'))} := \eta_{(p',(a',b'))}(f_{(p',(\emptytup,b'))})\tb
  \mfor (p',(\emptytup,b'))\in \dx\mand a':=A(\mr{p}{p'})(a).\]
\end{definition}

\begin{lemma}[Right Adjoint]\label{lem:mltt:radj}
$\spi{B}{}$ is right adjoint to $\pb{\grop{B}}{}$. Furthermore, for every natural transformation $g:A'\arr A$, we have the following coherence property
\[\pb{g}{(\spi{B}{C})}=\spi{\pb{g}{B}}{\pb{(\pbf{g}{B})}{C}}.\]
\end{lemma}
\input{proof_radj}

\begin{example}[Continuing Ex.~\ref{ex:mltt:mltt}]\label{ex:mltt:mltt2}
The $\Sign$-indexed set $\grot{\Con}{\Tp}$ maps every MLTT-signature $\Sigma$ to the set of pairs $(\Gamma,S)$ such that $\ofkind{\Sigma}{\Gamma}{S}{\kity}$. The projection $\grop{\Tp}$ is a natural transformation $\grot{\Con}{\Tp}\arr\Con$ such that $(\grop{\Tp})_\Sigma:(\Gamma,S)\mapsto \Gamma$.

We define $\Tm$ such that $\indset{\Sign,(\grot{\Con}{\Tp}),\Tm}$: The set $\Tm(\Sigma,(\Gamma,S))$ contains the terms $s$ such that $\oftype{\Sigma}{\Gamma}{s}{S}$. $\Tm((\Sigma,(\Gamma,S))\leq(\Sigma',(\Gamma,S)))$ is an inclusion.

Then we have $\indset{\Sign,\Con,\pgro{\Tp}{\Tm}}$, and $\pgro{\Tp}{\Tm}$ maps $(\Sigma,\Gamma)$ to the set of pairs $(S,s)$ such that $\oftype{\Sigma}{\Gamma}{s}{S}$.

To exemplify Def.~\ref{def:mltt:radj_aux}, fix an element $x=(\Sigma,\Gamma)\in\grodf{\Sign}{\Con}$. Then we have $\ix_{\Sigma'}(\es)=\Gamma$ for every $\Sigma\sq\Sigma'$. $\iswithout{\Tp}{x}$ maps the pair $(\Sigma',\es)$ where $\Sigma\sq\Sigma'$ to $\Tp(\Sigma',\ix_{\Sigma'}(\es))=\Tp(\Sigma',\Gamma)$. If $S\in\Tp(\Sigma',\ix_{\Sigma'}(\es))$, then $\iswithout{\Tm}{x}$ maps $(\Sigma',(\es,S))$ to the set $\Tm(\Sigma',(\ix_{\Sigma'}(\es),S))$.

Now we have $\indset{\Sign,\Con,\spi{\Tp}{\Tm}}$, and $\spi{\Tp}{\Tm}$ maps $(\Sigma,\Gamma)$ to the set of indexed elements of $\iswithout{\Tm}{x}$. Those are the families that assign to every $(\Sigma',(\es,S))$ a term $s_{(\Sigma',(\es,S))}\in\iswithout{\Tm}{x}(\Sigma',(\es,S))=\Tm(\Sigma',(\Gamma,S))$ such that $s_{(\Sigma',(\es,S))}=s_{(\Sigma'',(\es,S))}$ whenever $\Sigma'\sq\Sigma''$.
\end{example}

Above, we called $\STerms{C}$ the naive candidate for the right adjoint, and indeed the adjointness implies $\STerms{\spi{B}{C}}\isom \STerms{C}$. We define the isomorphisms explicitly because we will use them later on:

\begin{lemma}\label{lem:mltt:split}
Assume $\indset{P,A,B}$ and $\indset{P,(\grot{A}{B}),C}$. For $t\in\STerms{C}$ and $x:=(p,a)\in\grod{A}$, let $t^x\in\STerms{\Cx}$ be given by
\[(t^x)_{(p',(\emptytup,b'))}=t_{(p',(a',b'))}\tb\mwhere a':=A(\mr{p}{p'})(a).\]
And for $f\in\STerms{\spi{B}{C}}$ and $x:=(p,(a,b))\in\grod{\grot{A}{B}}$, we have $f_{(p,a)}\in\STerms{\Cx}$; thus, we can put
 \[f^{x}:=(f_{(p,a)})_{(p,(\emptytup,b))}\;\in\;C(p,(a,b)).\]
Then the sets 
$\STerms{C}$ and $\STerms{\spi{B}{C}}$ are in bijection via 
\[\STerms{C}\;\ni\; t\;\stackrel{\spl{A}{B}{C}{-}}{\longmapsto}\;(t^x)_{x\in\grod{A}}\;\in\;\STerms{\spi{B}{C}}\]
and
\[\STerms{\spi{B}{C}}\;\ni\; f \;\stackrel{\unspl{A}{B}{C}{-}}{\longmapsto}\;(f^{x})_{x\in\grod{\grot{A}{B}}}\;\in\;\STerms{C}.\]
\end{lemma}
\begin{proof}
This follows from the right adjointness by easy computations.
\myqed\end{proof}
Intuitively, $\spl{A}{B}{C}{t}$ turns $t\in\STerms{C}$ into a $\grod{A}$-indexed set by splitting it into components. And $\unspl{A}{B}{C}{f}$ amalgamates such a tuple of components back together. Syntactically, these operations correspond to currying and uncurrying, respectively.

Then we need one last notation. For $\indset{P,A}$, indexed elements $a\in\STerms{A}$ behave like mappings with domain $P$. We can precompose such indexed elements with fibrations $f:Q\arr P$ to obtain $Q$-indexed elements of $\STerms{\ö{f}{A}}$.
\begin{definition}\label{def:mltt:apply}
Assume $\indset{P,A}$, $f:Q\arr P$, and $a\in\STerms{A}$. $\atcomp{f}{a}\in\STerms{\ö{f}{A}}$ is defined by:
 $(\atcomp{f}{a})_q:=a_{f(q)}\mfor q\in Q$.
\end{definition}

%% file: proof_radj.tex
\begin{proof}
Assume $\indset{P,A,B}$, $\indset{P,\grot{A}{B},C}$, and $x=(p,a)\in\grod{A}$. Let $\yonex\in\Pres{\grod{A}}$ be the covariant representable functor of $x$ mapping $x'\in\grod{A}$ to a singleton iff $x\leq x'$ and to the empty set otherwise. Since we know the right adjoint exists, we can use the Yoneda lemma for covariant functors to derive sufficient and necessary constraints for $\spi{B}{}$ to be a right adjoint:
\[\mathll{(\spi{B}{C})(x)\isom \Hom_{\Pres{\grod{A}}}(\yonex,\spi{B}{C})
    \isom \Hom_{\Pres{\grod{\grot{A}{B}}}}(\pb{\grop{B}}{\yonex},C)   \nl
    \isom\Hom_{\Fib{\grod{\grot{A}{B}}}}(\groF{\pb{\grop{B}}{\yonex}},\groF{C}).}\]
Let $\ix$ be as in Def.~\ref{def:mltt:radj_aux}. Let $\Fibb{Q}$ be the category of (not necessarily canonical) fibrations on $Q$. Then it is easy to check that $\groF{\pbf{\ix}{B}}$ seen as a fibration with domain $\dx$ and $\groF{\pb{\grop{B}}{\yonex}}$ are isomorphic in $\Fibb{\grod{\grot{A}{B}}}$. (They are not isomorphic in $\Fib{\grod{B}}$ because the former is not canonical and thus not an object of $\Fib{\grod{B}}$.)
Using the fullness of $\Fib{Q}$, we obtain
\[\mathll{(\spi{B}{C})(x)\isom \Hom_{\Fibb{\grod{\grot{A}{B}}}}(\groF{\pbf{\ix}{B}},\groF{C}) \nl
  =\{f:\dx\arr\grod{C}\|\ö{f}{\groF{C}}=\groF{\pbf{\ix}{B}}\}.}\]
And using the definition of $\Cx$ as a pullback, we obtain \[(\spi{B}{C})(x)\isom\{f:\dx\arr\grod{\Cx}\|\ö{f}{\groF{\Cx}}=\id{\dx}\}\isom\STerms{\Cx}.\]
And this is indeed how $\spi{B}{C}$ is defined. The value of $\spi{B}{C}$ on morphisms is verified similarly.

To show the coherence property, we assume $\indset{P,A'}$, $g:A'\arr A$, and $x':=(p,a')\in \grod{A'}$.
We abbreviate as follows: $a:=g_p(a')$, $x:=(p,a)$, $B':=\pb{g}{B}$, and $C':=\pb{(\pbf{g}{B})}{C}$.
Furthermore, we write $\ixp$, $\Axp$, $\Bxp$, and $\Cxp$ according to Def.~\ref{def:mltt:radj_aux}. Note that $\Axp=\Ax$.


Now coherence requires $\pb{g}{\spi{B}{C}}=\spi{B'}{C'}$. And that follows if we show that
 \[\Bxp=\Bx  \tb\mand\tb  \Cxp=\Cx.\]
Using Lem.~\ref{lem:mltt:pullback}, this follows from $\ö{\ixp}{g}=\ix$, which is an equality between natural transformations from $\Ax=\Axp$ to $A$ in $\Pres{P}$.
And to verify the latter, assume $o\in P$. The maps $\ö{\ixp_{o}}{g_{o}}$ and $\ix_{o}$ have domain 
$\es$ or $\{\emptytup\}$. In the former case, there is nothing to prove. In the latter case, put \[a'_o:=\ixp_{o}(\emptytup)=A'(\mr{p}{o})(a')\tb\mand\tb a_o:=\ix_{o}(\emptytup)=A(\mr{p}{o})(a).\]
Then we need to show $g_{o}(a'_o)=a_o$. And that is indeed the case because of the naturality of $g$ as indicated in 
\begin{center}
\begin{tikzpicture}
\node (a') at (0,2)
    {$a'$};
\node (a'0) at (3,2)
    {$a'_o$};
\node (a) at (0,0)
    {$a$};
\node (a0) at (3,0)
    {$a_o$};
\draw[|-\arrowtip](a') -- node[above] {$A'(\mr{p}{o})$} (a'0);
\draw[|-\arrowtip](a') -- node[left] {$g_p$} (a);
\draw[|-\arrowtip](a'0) -- node[right] {$g_{o}$} (a0);
\draw[|-\arrowtip](a) -- node[above] {$A(\mr{p}{o})$} (a0);
\end{tikzpicture}
\end{center}
\myqed\end{proof}

%% file: semantics.tex
Using the LCC structure developed in Sect.~\ref{sec:mltt:functors}, the definition of the semantics is straightforward and well-known. To demonstrate its simplicity, we spell it out in an elementary way. The semantics is defined by induction on the derivations of the judgments listed in Fig.~\ref{fig:mltt:judgments}.

Firstly, for every signature $\issig{\Sigma}$, we define models $I$, which provide interpretations $\sem{c}{I}$ and $\sem{a}{I}$ for all symbols declared in $\Sigma$. The models are Kripke-models, i.e., a $\Sigma$-model $I$ is based on a poset $\bps{I}$ of worlds.

Secondly, $I$ extends to an interpretation function $\sem{-}{I}$, which interprets all $\Sigma$-expressions. We will omit the index $I$ if no confusion is possible.
\donotprintmodel
$\sem{-}{I}$ is such that
\begin{itemize}
	\item if $\iscont{\Sigma}{\Gamma}$, then $\sem{\Gamma}{I}$ is a poset (which has a canonical projection to $P$),
	\item if $\issubs{\Sigma}{\Gamma}{\Gamma'}{\gamma}$, then $\sem{\gamma}{I}:\sem{\Gamma'}{I}\arr\sem{\Gamma}{I}$ is a monotone function,
  \item if $\ofkind{\Sigma}{\Gamma}{S}{\kity}$, then $\semc{\Gamma}{S}{I}$ is an indexed set on $\sem{\Gamma}{I}$,
  \item if $\oftype{\Sigma}{\Gamma}{s}{S}$, then $\semc{\Gamma}{s}{I}$ is an indexed element of $\semc{\Gamma}{S}{I}$.
\end{itemize}

Thirdly, the judgments $\isequal{\Sigma}{\Gamma}{S}{S'}$ and $\isequal{\Sigma}{\Gamma}{s}{s'}$ correspond to a soundness result, which we will prove in Sect.~\ref{sec:mltt:soundness}.
\medskip

The poset $\bps{I}$ of worlds plays the same role as the various posets $\sem{\Gamma}{}$ --- it interprets the empty context. In this way, $P$ can be regarded as interpreting an implicit or relative context. This is in keeping with the practice of type theory (and category theory), according to which closed expressions may be considered relative to some fixed but unspecified context (respectively, base category).

For a typed term $\oftype{\Sigma}{\Gamma}{s}{S}$, both $\semc{\Gamma}{s}{I}$ and $\semc{\Gamma}{S}{I}$ are indexed over $\sem{\Gamma}{I}$. If $\Gamma=\decl{x_1}{S_1},\ldots,\decl{x_n}{S_n}$, an element of $\sem{\Gamma}{I}$ has the form $(p,(a_1,\ldots,a_n))$ where $p\in\bps{I}$ and $a_i\in\semc{\decl{x_1}{S_1},\ldots,\decl{x_{i-1}}{S_{i-1}}}{S_i}{I}(p,(a_1,\ldots,a_{i-1}))$.
Intuitively, $a_i$ is an assignment to the variable $x_i$ in world $p$.
And if an assignment $(p,\ass)$ is given, the interpretations of $s$ and $S$ satisfy $\semcp{\Gamma}{s}{I}{(p,\ass)}\in\semc{\Gamma}{S}{I}(p,\ass)$. This is illustrated in the left diagram in Fig.~\ref{fig:mltt:semanticstermstypes}.

If $\gamma$ is a substitution $\Gamma\arr\Gamma'$, then $\sem{\gamma}{I}$ maps assignments $(p,\ass')\in\sem{\Gamma'}{I}$ to assignments $(p,\ass)\in\sem{\Gamma}{I}$. And a substitution in types and terms is interpreted by pullback, i.e., composition. This is illustrated in the right diagram in Fig.~\ref{fig:mltt:semanticssubstitution}, whose commutativity expresses the coherence. We will state this more precisely in Sect.~\ref{sec:mltt:substitution}. 


Sum types are interpreted naturally as the dependent sum of indexed sets given by the left adjoint. And pairing and projections have their natural semantics. Product types are interpreted as exponentials using the right adjoint. A $\lambda$-abstraction $\lam{x}{S}t$ is interpreted by first interpreting $t$ and then splitting it as in Lem.~\ref{lem:mltt:split}. And an application $f\;s$ is interpreted by amalgamating the interpretation of $f$ as in Lem.~\ref{lem:mltt:split} and using the composition from Def.~\ref{def:mltt:apply}.

\begin{figure}[tbh]
\begin{center}
\begin{tikzpicture}
\node (S) at (0,4) {$\grod{\semc{\Gamma}{S}{I}}$};
\node (G1) at (-2.5,2) {$\sem{\Gamma}{I}$};
\node (G2) at (0,2) {$\sem{\Gamma}{I}$};
\draw[-\arrowtip](S) -- node[right] {$\groF{\semc{\Gamma}{S}{I}}$} (G2);
\draw[-\arrowtip](G1) -- node[left] {$\groF{\semc{\Gamma}{s}{I}}$} (S);
\draw[-\arrowtip](G1) -- node[above] {$\id{}$} (G2);
\end{tikzpicture}
\begin{tikzpicture}
\node (G2) at (0,2) {$\sem{\Gamma}{I}$};
\node (G'2) at (-3,2) {$\sem{\Gamma'}{I}$};
\node (St) at (-1.5,0) {$\Set$};
\draw[-\arrowtip](G'2) -- node[above] {$\sem{\gamma}{I}$} (G2);
\draw[-\arrowtip](G2) -- node[right] {$\semc{\Gamma}{S}{I}$} (St);
\draw[-\arrowtip](G'2) -- node[left] {$\semc{\Gamma'}{\subap{\gamma}{S}}{I}$} (St);
\end{tikzpicture}
\end{center}
\caption{Semantics of Terms, Types, and Substitution}\label{fig:mltt:semanticstermstypes}\label{fig:mltt:semanticssubstitution}
\end{figure}

\printmodel
\begin{definition}[Models]\label{def:mltt:models}
For a signature $\Sigma$, $\Sigma$-\defemph{models} are defined as follows:
\begin{itemize}
 \item A model $I$ for the empty signature $\cdot$ is a poset $\bps{I}$.
 \item A model $I$ for the signature $\Sigma,\decl{c}{S}$ consists of
   a $\Sigma$-model $I_{\Sigma}$ and an indexed element $\sem{c}{I}\in\STerms{\semc{\cdot}{S}{I_{\Sigma}}}$.
 \item A model $I$ for the signature $\Sigma,\declk{a}{\Gammak}$ consists of
   a $\Sigma$-model $I_{\Sigma}$ and an indexed set $\sem{a}{I}$ over $\sem{\Gammak}{I_{\Sigma}}$.
\end{itemize}
\end{definition}
\donotprintmodel

\begin{definition}[Model Extension]\label{def:mltt:extension}
\newcommand{\assl}{(p,\ass)}
\newcommand{\GammaS}{\Gamma,\decl{x}{S}}
The extension of a model is defined by induction on the typing derivations. Therefore, we can assume in each case that all occurring expressions are well-formed. For example in the case for $\semc{\Gamma}{f\;s}{}$, $f$ has type $\P{x}{S}T$ and $s$ has type $S$.

\begin{itemize}
\item Contexts: The elements of the poset $\sem{\decl{x_1}{S_1},\ldots,\decl{x_n}{S_n}}{}$ are the tuples $(p,(a_1,\ldots,a_n))$ such that
\[\mathll{
  p\in \bps{}\\
  a_1\in \semc{\cdot}{S_1}{}(p,\emptytup)\\
  \vdots\\
  a_n\in \semc{\decl{x_1}{S_1},\ldots,\decl{x_{n-1}}{S_{n-1}}}{S_n}{}(p,(a_1,\ldots,a_{n-1}))
}\]
In particular $\sem{\cdot}{}=\bps{}\times\{\es\}$.
The ordering of this poset is inherited from the $n$-times iterated category of elements, to which it is canonically isomorphic. The first projection from $\sem{\Gamma}{}$ is a canonical fibration, and we write $\groI{\sem{\Gamma}{}}$ for the corresponding indexed set.

\item Substitutions $\gamma=\sb{x_1}{s_1,\ldots,\sb{x_n}{s_n}}$ from $\Gamma$ to $\Gamma'$:
  \[\sem{\gamma}{}:(p,\ass')\mapsto \big(p,(\semcp{\Gamma'}{s_1}{}{(p,\ass')},\ldots,\semcp{\Gamma'}{s_n}{}{(p,\ass')})\big)
                                    \tb\mfor (p,\ass')\in\sem{\Gamma'}{}\]
  We write $\groI{\sem{\gamma}{}}$ for the induced natural transformation $\groI{\sem{\Gamma'}{}}\arr\groI{\sem{\Gamma}{}}$.
  
  \item Basic types:
    \[\semc{\Gamma}{\appt{a}{\gammak}}{} := \ö{\sem{\gammak}{}}{\sem{a}{}}\]
\item Complex types:
  \begin{myeqnarray}[:=]
    \semc{\Gamma}{\unit}{}(p,\ass)        & \{\SunitE\} \\[.2cm]
    \semc{\Gamma}{\ident{s}{s'}}{}(p,\ass)&
       \cas{\{\SunitE\}\mifc \semcp{\Gamma}{s}{}{\assl}=\semcp{\Gamma}{s'}{}{\assl} \\ \es\mothw} \\[.2cm]
    \semc{\Gamma}{\S{x}{S}T}{}            & \pgro{\semc{\Gamma}{S}{}}{\semc{\GammaS}{T}{}} \\[.2cm]
    \semc{\Gamma}{\P{x}{S}T}{}            & \spi{\semc{\Gamma}{S}{}}{\semc{\GammaS}{T}{}}
  \end{myeqnarray}
  $\semc{\Gamma}{\unit}{}$ and $\semc{\Gamma}{\ident{s}{s'}}{}$ are only specified for objects; their extension to morphisms is uniquely determined.
\item Basic terms:
  \[\semcp{\Gamma}{c}{}{(p,\ass)} := \semp{c}{}{p}, \tb\tb
    \semcp{\decl{x_1}{S_1},\ldots,\decl{x_n}{S_n}}{x_i}{}{(p,(a_1,\ldots,a_n))} := a_i\]
\item Complex terms:
  \begin{myeqnarray}[:=]
    \semcp{\Gamma}{\unitE}{}{\assl}       & \SunitE  \\
    \semcp{\Gamma}{\refl{s}}{}{\assl}     & \SunitE \\
    \semcp{\Gamma}{\pair{s}{s'}}{}{\assl} & (\semcp{\Gamma}{s}{}{\assl},\semcp{\Gamma}{s'}{}{\assl}) \\
    \semcp{\Gamma}{\pi_i(u)}{}{\assl}     & a_i \tb\mwhere\semcp{\Gamma}{u}{}{\assl}=(a_1,a_2) \\
    \semc{\Gamma}{\lam{x}{S}t}{}          & \spl{\sem{\Gamma}{}}{\semc{\Gamma}{S}{}}{\semc{\GammaS}{T}{}}{\semc{\GammaS}{t}{}} \\
    \semc{\Gamma}{f\;s}{}                & \atcomp{(\ö{\groF{\semc{\Gamma}{s}{}}}{\assoc})}
                                                  {\unspl{\sem{\Gamma}{}} {\semc{\Gamma}{S}{}} {\semc{\GammaS}{T}{}} {\semc{\Gamma}{f}{}}} \\
  \end{myeqnarray}
  Here $\assoc$ maps $((p,\alpha),a)$ to $(p,(\alpha,a))$.
\end{itemize}
\end{definition}\vspace{2 pt}

\noindent Since the same expression may have more than one well-formedness derivation, the well-definedness of Def.~\ref{def:mltt:extension} must be proved in a joint induction with the proof of Thm.~\ref{thm:mltt:sound} below (see also \cite{lcccstreicher}). And because of the use of substitution, e.g., for application of function terms, the induction must be intertwined with the proof of Thm.~\ref{thm:mltt:substitution} as well.

\begin{example}[Continuing Ex.~\ref{ex:mltt:cat}]\label{ex:mltt:cat2}
A model of the signature $\excat$ over an indexing poset $P$ is the same thing as a functor from $P$ into $\Cat$, the category of (small) categories. In more detail, assume a poset $P$ and a functor $F:P\arr\Cat$. Then we obtain a model of the signature $\excat$ as follows:
\begin{itemize}
	\item The underlying poset is $P$.
	\item $\sem{\exob}{\exmod}$ is the indexed set over $P$ mapping
	  \begin{itemize}
	    \item every $p\in P$ to the set of objects of $F(p)$,
	    \item every morphism $\mr{p}{p'}$ to the object-part of $F(\mr{p}{p'})$.
    \end{itemize}
  \item $\sem{\decl{x}{\exob},\decl{y}{\exob}}{\exmod}$ is a poset containing tuples $(p,(a,b))$ for $a,b\in F(p)$. We obtain $(p,(a,b))\leq (p',(a',b'))$ iff $p\leq p'$ and $a'=F(p\leq p')(a)$ and $b'=F(p\leq p')(b)$.
    Then $\sem{\exmor}{\exmod}$ is the indexed set over $\sem{\decl{x}{\exob},\decl{y}{\exob}}{\exmod}$ mapping
   \begin{itemize}
	   \item every $(p,(a,b))$ to the set $\Hom_{F(p)}(a,b)$,
	   \item every $(p,(a,b))\leq (p',(a',b'))$ to the morphism part of $F(\mr{p}{p'})$ restricted to a map from $\Hom_{F(p)}(a,b)$ to $\Hom_{F(p')}(a',b')$.
   \end{itemize}
  \item Next we define $\sem{\exid}{\exmod}\in\STerms{\semc{\cdot}{\P{x}{\exob}\exmor\;x\;x}{\exmod}}$ as $\spl{}{}{}{e}$ (using Lem.~\ref{lem:mltt:split}) where $e\in \STerms{\semc{\decl{x}{\exob}}{\exmor\;x\;x}{\exmod}}$ is defined as follows. $\semc{\decl{x}{\exob}}{\exmor\;x\;x}{\exmod}$ maps $(p,a)$ for $a\in \semc{\cdot}{\exob}{\exmod}(p)$ to the set $\Hom_{F(p)}(a,a)$, and we put $e_{(p,a)}:=\id{a}$. \\
  Because $F$ is a functor, we have
   \[\semc{\decl{x}{\exob}}{\exmor\;x\;x}{\exmod}((p,a)\leq (p',a'))(\id{a})=\id{a'}.\]
  Therefore, $e$ is indeed an indexed element.
  \item $\excomp$ is interpreted as composition in $F(p)$ in the same manner as $\exid$ applying Lem.~\ref{lem:mltt:split} five times.
  \item The interpretations of the constants representing axioms such as $\exneutr$ are uniquely determined. And they exist because all $F(p)$ are categories.
\end{itemize}
\end{example}


%% file: substitution.tex
Parallel to Lem.~\ref{lem:mltt:substitutionapply}, we obtain the following central result about the semantics of substitutions. It expresses the coherence of our models.

\begin{theorem}[Substitution]\label{thm:mltt:substitution}
Assume $\issubs{\Sigma}{\Gamma}{\Gamma'}{\gamma}$. Then:
\begin{center}
\begin{tabular}{@{if\tb}l@{\tb then\tb}l}
	$\issubs{\Sigma}{\Delta}{\Gamma}{\delta}$ & $\sem{\ö{\delta}{\gamma}}{}=\ö{\sem{\gamma}{}}{\sem{\delta}{}}$, \\
	$\ofkind{\Sigma}{\Gamma}{S}{\kity}$       & $\semc{\Gamma'}{\subap{\gamma}{S}}{}=\ö{\sem{\gamma}{}}{\semc{\Gamma}{S}{}}$,  \\
	$\oftype{\Sigma}{\Gamma}{s}{S}$           &
	            $\semc{\Gamma'}{\subap{\gamma}{s}}{}=\atcomp{\sem{\gamma}{}}{\semc{\Gamma}{s}{}}$.
\end{tabular}
\end{center}
\end{theorem}\bigskip

\noindent Before we give the proof of Thm.~\ref{thm:mltt:substitution}, we establish some auxiliary results:
\begin{lemma}\label{lem:mltt:substitutionaux}
Assume $\issubs{\Sigma}{\Gamma}{\Gamma'}{\gamma}$ and $\oftype{\Sigma}{\Gamma}{S}{\kity}$ and thus also
 \[\issubs{\Sigma}{\;\Gamma,\decl{x}{S}\;}{\;\Gamma',\decl{x}{\subap{\gamma}{S}}\;}{\;\gamma,\sb{x}{x}\;}.\]
 Furthermore, assume the induction hypothesis of Thm.~\ref{thm:mltt:substitution} for the involved expressions.
 Then we have:
  \[\sem{\gamma,\sb{x}{x}}{}
   =\groF{\pbf{\groI{\sem{\gamma}{}}}{\semc{\Gamma}{S}{}}}.\]
\end{lemma}
\begin{proof}
This follows by direct computation.
\myqed\end{proof}

\begin{lemma}\label{lem:mltt:splitcoherence}
Assume $\indset{P,A,B}$, $\indset{P,\grot{A}{B},C}$, $\indset{P,A'}$, a natural transformation $g:A'\arr A$, and $t\in\STerms{C}$. Then for $x'\in\grod{A'}$:
 \[\spl{A}{B}{C}{\atcomp{\groF{\pbf{g}{B}}}{t}}_{x'}=\spl{A'}{\pb{g}{B}}{\pb{(\pbf{g}{B})}{C}}{t}_{\groF{g}(x')}.\]
\end{lemma}
\begin{proof}
This follows by direct computation.
\myqed\end{proof}

\begin{proof}[Proof of Thm.~\ref{thm:mltt:substitution}]
The proofs of all subtheorems are intertwined in an induction on the typing derivations; in addition, the induction is intertwined with the proof of Thm.~\ref{thm:mltt:sound}.

The case of an empty \emph{substitution} $\delta$ is trivial. For the remaining cases, assume $\delta=\sb{x_1}{s_1},\ldots,\sb{x_n}{s_n}$ and $(p,\alpha')\in\sem{\Gamma'}{}$. Then applying the composition of substitutions, the semantics of substitutions, the induction hypothesis for terms, and the semantics of substitutions, respectively, yields:
  \[\mathll{
    \sem{\ö{\delta}{\gamma}}{}(p,\ass')
  = \sem{\sb{x_1}{\subap{\gamma}{s_1}},\ldots,\sb{x_n}{\subap{\gamma}{s_n}}}{}(p,\ass')
  = \big(p,\big(\semcp{\Gamma'}{\subap{\gamma}{s_i}}{}{(p,\ass')}\big)_{i=1,\ldots,n}\big) \nl
  = \big(p,\big(\semcp{\Gamma}{s_i}{}{\sem{\gamma}{}(p,\ass')}\big)_{i=1,\ldots,n}\big)
  = (\ö{\sem{\gamma}{}}{\sem{\delta}{}})(p,\ass')
  }\]

\noindent
The cases for \emph{types} are as follows:
\begin{itemize}
\item $\appt{a}{\gammak}$:
  Using the definition of substitution and the semantics of application, we obtain:
  \[\semc{\Gamma'}{\subap{\gamma}{\appt{a}{\gammak}}}{}
  = \semc{\Gamma'}{\appt{a}{(\ö{\gamma}{\gammak})}}{}
  = \ö{\sem{\ö{\gamma}{\gammak}}{}}{\sem{a}{}}\]
  And similarly we obtain:
  \[\semc{\Gamma}{\appt{a}{\gammak}}{}
  = \ö{\sem{\gammak}{}}{\sem{a}{}}\]
  Then the needed equality follows from the induction hypothesis for $\gammak$.
\begin{center}
\begin{tikzpicture}
\node (G') at (0,0)
    {$\sem{\Gamma'}{}$};
\node (G0) at (2,2)
    {$\sem{\Gammak}{}$};
\node (G) at (2,0)
    {$\sem{\Gamma}{}$};
\node (S) at (2,4)
    {$\Set$};
\draw[-\arrowtip](G') -- node[above] {$\sem{\gamma}{}$} (G);
\draw[-\arrowtip](G') -- node[left] {$\ö{\sem{\gamma}{}}{\sem{\gammak}{}}$} (G0);
\draw[-\arrowtip](G) -- node[right] {$\sem{\gammak}{}$} (G0);
\draw[-\arrowtip](G0) -- node[right] {$\sem{a}{}$} (S);
\end{tikzpicture}
\end{center}
\item $\unit$: Trivial.

\item $\ident{s}{s'}$: This follows directly from the induction hypothesis for $s$ and $s'$.

\item $\S{x}{S}T$: This follows directly by combining the induction hypothesis as well as Lem.~\ref{lem:mltt:ladj} and~\ref{lem:mltt:substitutionaux}.

\item $\P{x}{S}T$: This follows directly by combining the induction hypothesis as well as Lem.~\ref{lem:mltt:radj} and~\ref{lem:mltt:substitutionaux}.
\end{itemize}

\noindent
For the cases of a \emph{term} $s$, let us assume a fixed $(p,\ass')\in\sem{\Gamma'}{}$ and $(p,\ass):=\sem{\gamma}{}(p,\ass')$.
Then we need to show
 \[\semcp{\Gamma'}{\subap{\gamma}{s}}{}{(p,a')} = \semcp{\Gamma}{s}{}{(p,a)}.\]
\begin{itemize}
\item $c$: Clear because $\subap{\gamma}{c}=c$. 

\item $x$: Assume $x$ occurs in position $i$ in $\Gamma$, and let $\sb{x}{s}$ be in $\gamma$. Further, assume $\ass'=(a'_1,\ldots,a'_n)$ and $\ass=(a_1,\ldots,a_n)$. Then by the properties of substitutions: $\semcp{\Gamma'}{\subap{\gamma}{x}}{}{(p,\ass')}=\semcp{\Gamma'}{s}{}{(p,\ass')}=a_i$. And that is equal to $\semcp{\Gamma}{x}{}{(p,\ass)}$.

\item $\refl{s}$: Trivial.

\item $\unitE$: Trivial.

\item $\pair{s}{s'}$: Because $\subap{\gamma}{\pair{s}{s'}}=\pair{\subap{\gamma}{s}}{\subap{\gamma}{s'}}$, this case follows immediately from the induction hypothesis.

\item $\pi_i(u)$ for $i=1,2$: Because $\subap{\gamma}{\pi_i(s)}=\pi_i(\subap{\gamma}{s})$, this case follows immediately from the induction hypothesis.

\item $\lam{x}{S}t$: By the definition of substitution, the semantics of $\lambda$-abstraction, the induction hypothesis, and Lem.~\ref{lem:mltt:substitutionaux}, respectively, we obtain:
  \[\mathll[c]{
    \semc{\Gamma'}{\subap{\gamma}{\lam{x}{S}t}}{}
  = \semc{\Gamma'}{\lam{x}{\subap{\gamma}{S}}\subapx{\gamma}{x}{t}}{}
  = \spl{}{}{}{{\semc{\Gamma',\decl{x}{\subap{\gamma}{S}}}{\subapx{\gamma}{x}{t}}{}}} \\[0.2cm]
  = \spl{}{}{}{{\atcomp{\sem{\gamma,\sb{x}{x}}{}}{\semc{\Gamma,\decl{x}{S}}{t}{}}}} \\[0.2cm]
  = \spl{}{}{}{\atcomp
    {\groF{\pbf{\groI{\sem{\gamma}{}}}{\semc{\Gamma}{S}{}}}}
    {{\semc{\Gamma,\decl{x}{S}}{t}{}}}
    }.
  }\]
  Furthermore, we have $\semc{\Gamma}{\lam{x}{S}t}{}=\spl{}{}{}{{\semc{\Gamma,\decl{x}{S}}{t}{}}}$. Then the result follows by using Lem.~\ref{lem:mltt:splitcoherence} and $\groF{\groI{\sem{\gamma}{}}}=\sem{\gamma}{}$.
  
\item $f\;s$: We evaluate both sides of the needed equation. Firstly, on the left-hand side, we obtain by the definition of substitution, the semantics of application, and the induction hypothesis, respectively:
 \[\mathll[c]{
   \semc{\Gamma'}{\subap{\gamma}{f\;s}}{}
 = \semc{\Gamma'}{\subap{\gamma}{f}\;\subap{\gamma}{s}}{}
 = \atcomp{(\ö{\groF{\semc{\Gamma'}{\subap{\gamma}{s}}{}}}{\assoc})}{\unspl{}{}{}{\semc{\Gamma'}{\subap{\gamma}{f}}{}}} \nl
 = \atcomp{(\ö{\groF{\atcomp{\sem{\gamma}{}}{\semc{\Gamma}{s}{}}}}{\assoc})} {\unspl{}{}{}{\atcomp{\sem{\gamma}{}}{\semc{\Gamma}{f}{}}}}.
 }\]
 To compute the value at $(p,\ass')$ of this indexed element, we first compute $(\atcomp{\sem{\gamma}{}}{\semc{\Gamma}{s}{}})_{(p,\ass')}$, say we obtain $b$. Then we can compute $\unspl{}{}{}{\atcomp{\sem{\gamma}{}}{\semc{\Gamma}{f}{}}}_{(p,(\ass',b))}$. Using the notation from Lem.~\ref{lem:mltt:split}, the left-hand side evaluates to
 \[\big(\atcomp{\sem{\gamma}{}}{\semc{\Gamma}{f}{}}\big)^{(p,(\ass',b))}=
   (\semc{\Gamma}{f}{}_{(p,\ass)})_{(p,(\emptytup,b))}.\]
Secondly, on the right-hand side, we have by the semantics of application:
 \[\semc{\Gamma}{f\;s}{}
 = \atcomp{(\ö{\groF{\semc{\Gamma}{s}{}}}{\assoc})}{\unspl{}{}{}{\semc{\Gamma}{f}{}}}.\]
 When computing the value at $(p,\ass)$ of this indexed element, we obtain in a first step $\unspl{}{}{}{\semc{\Gamma}{f}{}}_{(p,(\ass,b))}$. And evaluating further, this yields $(\semc{\Gamma}{f}{}_{(p,\ass)})_{(p,(\emptytup,b))}$.

Thus, the equality holds as needed.
\myqed
\end{itemize}
\end{proof}

%% file: soundness.tex
We have already mentioned the soundness result, which states that the interpretation takes the syntactic judgments for equality of terms and types to corresponding semantic judgments:

\begin{theorem}[Soundness]\label{thm:mltt:sound}
Assume a signature $\Sigma$, and a context $\Gamma$. If $\isequal{\Sigma}{\Gamma}{S}{S'}$ for two well-formed types $S$, $S'$, then in every $\Sigma$-model: \[\semc{\Gamma}{S}{I}=\semc{\Gamma}{S'}{I}\in\Pres{\sem{\Gamma}{I}}.\]
And if $\isequal{\Sigma}{\Gamma}{s}{s'}$ for two well-formed terms $s$, $s'$ of type $S$, then in every $\Sigma$-model:
\[\semc{\Gamma}{s}{I}=\semc{\Gamma}{s'}{I}\in\STerms{\semc{\Gamma}{S}{I}}.\]
\end{theorem}

\begin{proof}
The soundness is proved by induction over all derivations; the induction is intertwined with the proof of Thm.~\ref{thm:mltt:substitution}. An instructive example is the rule $e_{\op{typing}}$. Its soundness states the following: If $\semc{\Gamma}{s}{}\in\STerms{\semc{\Gamma}{S}{}}$ and $\semc{\Gamma}{s}{}=\semc{\Gamma}{s'}{}$ and $\semc{\Gamma}{S}{}=\semc{\Gamma}{S'}{}$, then also $\semc{\Gamma}{s'}{}\in\STerms{\semc{\Gamma}{S'}{}}$. And this clearly holds.

Among the remaining rules for terms, the soundness of some rules is an immediate consequence of the semantics. These are: all rules from Fig.~\ref{fig:mltt:terms} except for $t_\lambda$ and $t_{\op{app}}$, and from Fig.~\ref{fig:mltt:conversions} the rules $e_{\ident{-}{-}}$, $e_{\op{id-uniq}}$, $e_\unitE$, $e_{\pair{-}{-}}$, $e_{\pi_1}$, $e_{\pi_2}$, and $e_{\op{app}}$.
\bigskip

The soundness of the rules $t_\lambda$ and $t_{\op{app}}$ follows by applying the semantics and Lem.~\ref{lem:mltt:split}. That leaves the rules $e_{\beta}$ and $e_{\op{funcext}}$, the soundness of which we will prove in detail.
\bigskip

For $e_{\beta}$, we interpret $(\lam{x}{S}t)\;s$ by applying the definition:
\[\mathll{
  \semc{\Gamma}{(\lam{x}{S}t)\;s}{}
 =\atcomp{(\ö{\groF{\semc{\Gamma}{s}{}}}{\assoc})}{\unspl{}{}{}{\semc{\Gamma}{\lam{x}{S}t}{}}} \nl
 =\atcomp{(\ö{\groF{\semc{\Gamma}{s}{}}}{\assoc})}{\unspl{}{}{}{\spl{}{}{}{\semc{\Gamma,\decl{x}{S}}{t}{}}}}
}\]
$\unspl{}{}{}{\spl{}{}{}{\semc{\Gamma,\decl{x}{S}}{t}{}}}$ is equal to $\semc{\Gamma,\decl{x}{S}}{t}{}$ by Lem.~\ref{lem:mltt:split}. Furthermore, we have $\subapo{t}{x}{s}=\subap{\gamma}{t}$ where $\gamma=\id{\Gamma},\sb{x}{s}$ is a substitution from $\Gamma,\decl{x}{S}$ to $\Gamma$. And interpreting $\gamma$ yields $\sem{\gamma}{}(p,\alpha)=(p,(\alpha,\semc{\Gamma}{s}{}_{(p,\alpha)}))$, i.e., $\sem{\gamma}{}=\ö{\groF{\semc{\Gamma}{s}{}}}{\assoc}$. Therefore, using Thm.~\ref{thm:mltt:substitution} for terms yields
\[\semc{\Gamma}{\subapo{t}{x}{s}}{}=\atcomp{(\ö{\groF{\semc{\Gamma}{s}{}}}{\assoc})}{\semc{\Gamma,\decl{x}{S}}{t}{}},\]
which concludes the soundness proof for $e_{\beta}$.
\bigskip

To understand the soundness of $e_{\op{funcext}}$, let us look at the interpretations of $f$ in the contexts $\Gamma$ and $\Gamma,\decl{y}{S}$:
\[\mathll{
  \unspl{}{}{}{\semc{\Gamma}{f}{}}\in\STerms{{\semc{\Gamma,\decl{x}{S}}{T}{}}},\tb
  \unspl{}{}{}{\semc{\Gamma,\decl{y}{S}}{f}{}}\in\STerms{{\semc{\Gamma,\decl{y}{S},\decl{x}{S}}{T}{}}}.
}\]
Let $\gamma$ be the inclusion substitution from $\Gamma$ to $\Gamma,\decl{y}{S}$. Then $\sem{\gamma}{}$ is the projection $\sem{\Gamma,\decl{y}{S}}{}\arr\sem{\Gamma}{}$ mapping elements $(p,(\alpha,a))$ to $(p,\alpha)$. Applying Thm.~\ref{thm:mltt:substitution} yields for arbitrary  $(p,\alpha)\in\sem{\Gamma}{}$ and $a',a\in\semc{\Gamma}{S}{}(p,\alpha)$:
 \[\unspl{}{}{}{\semc{\Gamma,\decl{y}{S}}{f}{}}_{(p,(\alpha,a',a))}
 = \unspl{}{}{}{\semc{\Gamma}{f}{}}_{(p,(\alpha,a))}.\]
And we have
 \[\semcp{\Gamma,\decl{y}{S}}{y}{}{(p,(\alpha,a'))}=a',\tb\mand\tb \groF{\semc{\Gamma,\decl{y}{S}}{y}{}}(p,(\alpha,a'))=(p,(\alpha,a'),a').\]
Putting these together yields
\[\mathll{
   \semc{\Gamma,\decl{y}{S}}{f\;y}{}_{(p,(\alpha,a'))}
 = \big(
     \atcomp{(\ö{\groF{\semc{\Gamma,\decl{y}{S}}{y}{}}}{\assoc})}{\unspl{}{}{}{\semc{\Gamma,\decl{y}{S}}{f}{}}}
   \big)_{(p,(\alpha,a'))} \nl
 = \unspl{}{}{}{\semc{\Gamma,\decl{y}{S}}{f}{}}_{(p,(\alpha,a',a'))}
 = \unspl{}{}{}{\semc{\Gamma}{f}{}}_{(p,(\alpha,a'))}
}\]
Therefore, the induction hypothesis applied to $\isequal{\Sigma}{\Gamma,\decl{y}{S}}{f\;y}{f'\;y}$ yields
\[\unspl{}{}{}{\semc{\Gamma}{f}{}} = \unspl{}{}{}{\semc{\Gamma}{f'}{}}.\]
And then Lem.~\ref{lem:mltt:split} yields
\[\semc{\Gamma}{f}{} = \semc{\Gamma}{f'}{}\]
concluding the soundness proof for $e_{\op{funcext}}$.
\bigskip

Regarding the rules for types in Fig.~\ref{fig:mltt:types} and~Fig.~\ref{fig:mltt:equalitytypes}, the soundness proofs are straightforward.
\myqed\end{proof}

%% file: completeness.tex
\printmodel

According to the propositions-as-types interpretation --- also known as the Curry-Howard correspondence --- a type $S$ holds in a model if its interpretation $\sem{S}{}$ is inhabited, i.e., the indexed set $\sem{S}{}$ has an indexed element. A type is valid if it holds in all models. Then soundness implies: If there is a term $s$ of type $S$ in context $\Gamma$, then in every $\Sigma$-model there is an indexed element of $\semc{\Gamma}{S}{}$, namely $\semc{\Gamma}{s}{}$. The converse is completeness: A type that has an indexed element in every model is inhabited.  Observe that the presence of (extensional) identity types then implies also the completeness of the equational term calculus because two terms are equal iff the corresponding identity type is inhabited.

The basic idea of the proof of completeness is to build the syntactic category, and then to construct a model out of it using categorical embedding theorems.

\begin{definition}\label{def:mltt:lccembedding}
A functor $F:\C\arr \mathcal{D}$ is called LCC if $\C$ is LCC and if $F$ preserves that structure, i.e., $F$ maps terminal object, products and exponentials in all slices $\slii{\C}{A}$ to corresponding structures in $\slii{\mathcal{D}}{F(A)}$. An LCC functor is called an LCC embedding if it is injective on objects, full, and faithful.
\end{definition}

We make use of a theorem from topos theory due to Butz and Moerdijk (\cite{spatialcover}) to establish the following central lemma.

\begin{lemma}\label{lem:mltt:lccembedding}
For every LCC category $\C$, there is a poset $P$ and an LCC embedding $E:\C\arr \Pres{P}$.
\end{lemma}
\begin{proof}
Clearly, the composition of LCC embeddings is an LCC embedding. We obtain $E:\C\arr\Pres{P}$ as a composite $\ö{E_1}{\ö{E_2}{E_3}}$.
Here $E_1:\C\arr \Presh{\C}$ is the Yoneda embedding, which maps $A\in\obj{\C}$ to $\Hom(-,A)$. This is well-known to be an LCC embedding.
$E_2$ maps a presheaf on $\C$ to a sheaf on a topological space $S$. $E_2$ is the inverse image part of the spatial cover of the topos $\Presh{\C}$ of presheaves on $\C$. This construction rests on a general topos-theoretical result established in \cite{spatialcover}, and we refer to \cite{awodey00topological} for the details of the construction of $S$, the definition of $E_2$, and the proof that $E_2$ is an LCC embedding.
Finally $E_3:\op{sh}(S)\arr \Presh{O(S)}$ includes a sheaf on $S$ into the category of presheaves on the poset $O(S)$ of open sets of $S$. That $E_3$ is an LCC embedding, can be verified directly. Finally, we put $P:=\catop{O(S)}$ so that $E$ becomes an LCC embedding into $\Pres{P}$.
\myqed\end{proof}

\begin{definition}[Term-Generated]
A $\Sigma$-model $I$ is called term-generated if for all closed $\Sigma$-types $S$ and every indexed element $e\in\STerms{\semc{\cdot}{S}{I}}$, there is a $\Sigma$-term $s$ of type $S$ such that $\semc{\cdot}{s}{I}=e$.
\end{definition}

\begin{theorem}[Model Existence]\label{thm:mltt:modelexistence}
For every signature $\Sigma$, there is a term-genera\-ted model $I$ such that for all types $\ofkind{\Sigma}{\Gamma}{S}{\kity}$
 \begin{equation}\label{eq:comp:types}
   \STerms{\semc{\Gamma}{S}{I}}\neq\es \tb\miff\tb \oftype{\Sigma}{\Gamma}{s}{S} \mfor\msome s,
  \end{equation}
and for all such terms $\oftype{\Sigma}{\Gamma}{s}{S}$ and $\oftype{\Sigma}{\Gamma}{s'}{S}$
 \begin{equation}\label{eq:comp:terms}
   \semc{\Gamma}{s}{I}=\semc{\Gamma}{s'}{I} \tb\miff\tb \isequal{\Sigma}{\Gamma}{s}{s'}.
 \end{equation}
\end{theorem}
\begin{proof}
It is well known how to construct the syntactic category $\C$ from $\Sigma$ and $\Gamma$ (\cite{lcccseely}). The objects of $\C$ are given by the set of all types $S$ such that $\ofkind{\Sigma}{}{S}{\kity}$ modulo the equivalence relation $\isequal{\Sigma}{}{S}{S'}$. We will write $[S]$ for the equivalence class of $S$.

The $\C$-morphisms from $[S]$ to $[S']$ are given by the terms $f$ such that $\oftype{\Sigma}{}{f}{S\arr S'}$ modulo the equivalence relation $\isequal{\Sigma}{}{f}{f'}$. We will write $[f]$ for the equivalence class of $f$.

It is straightforward to check that $\C$ is LCC (see, e.g., \cite{lcccseely}). For example, the exponential $f_2^{f_1}$ of two objects $\oftype{\Sigma}{}{f_1}{S_1\arr S}$ and $\oftype{\Sigma}{}{f_2}{S_2\arr S}$ in a slice $\slii{\C}{[S]}$ is given by
 \[\lam{u}{U}\pi_1(u)\tb\mwhere\tb U:=\S{x}{S}\big(\S{y_1}{S_1}\ident{x}{f_1\;y_1}\arr\S{y_2}{S_2}\ident{x}{f_2\;y_2}\big).\]
By Lem.~\ref{lem:mltt:lccembedding}, there are a poset $P$ and an LCC embedding $E:\C\arr\Pres{P}$. From those, we construct the needed model $I$ over $P$. Essentially, $I$ arises by interpreting every term or type as its image under $E$.

Firstly, assume a declaration $\decl{c}{S}$ in $\Sigma$. Since $\C$ only uses types and function terms, $E$ cannot in general be applied to $c$. But using the type $\unit$, every term $c$ of type $S$ can be seen as the function term $\lam{x}{\unit}c$ of type $\unit\arr S$. Therefore, we define $E'(c):=E([\lam{x}{\unit}c])$, which is an indexed element of $E([\unit\arr S])$. Since $\STerms{E([\unit\arr S])}$ and $\STerms{E([S])}$ are in bijection,  $E'(c)$ induces an indexed element of $E([S])$, which we use to define $\sem{c}{I}$.

Secondly, assume a declaration $\declk{a}{\Gammak}$ in $\Sigma$ for $\Gammak=x_1:S_1,\ldots,x_n:S_n$. $\sem{a}{I}$ must be an indexed set over $\sem{\Gammak}{I}$. For the same reason as above, $E$ cannot be applied directly to $a$. Instead, we use the type $U:=\S{x_1}{S_1}\ldots\S{x_n}{S_n}(\appt{a}{\id{\Gammak}})$. The fibration $\groF{E([U])}:\grodf{P}{E(U)}\arr P$ factors canonically through $\sem{\Gammak}{I}$, from which we obtain the needed indexed set $\sem{a}{I}$.

That $I$ is term-generated now follows directly from the fullness of $E$.  Finally, the required property (\ref{eq:comp:types}) clearly follows from $I$ being term-generated, and (\ref{eq:comp:terms}) from the fact that $E$ is faithful.
\myqed\end{proof}

The fact that the model $I$ just constructed is term-generated can be interpreted as functional completeness of the semantics: If a natural transformation of a certain type exists in every model, then it is syntactically definable. In more detail, let $I$ be the model constructed in Thm.~\ref{thm:mltt:modelexistence}, and assume a natural transformation $\eta:\semc{\cdot}{S}{I}\arr\semc{\cdot}{S'}{I}$ for some $\Sigma$-types $S$ and $S'$. Then there exists a $\Sigma$-term $f$ of type $S\arr S'$ such that $\eta$ arises from $\semc{\cdot}{f}{I}$ as follows. Put $\eta':=\unspl{P}{\semc{\cdot}{S}{I}}{\semc{\decl{x}}{S'}{I}}{\semc{\cdot}{f}{I}}\in\STerms{\semc{\decl{x}{S}}{S'}{I}}$. Then $\eta'$ maps pairs $(p,a)$ to elements of $\semc{\decl{x}{S}}{S'}{I}(p,a)=\semc{\cdot}{S'}{I}(p)$ for $a\in\semc{\cdot}{S}{I}(p)$. Then we obtain $\eta$ as $\eta_p:a\mapsto \eta'(p,a)$.

\begin{theorem}[Completeness]\label{thm:mltt:complete}
For every signature $\Sigma$ and any type $\ofkind{\Sigma}{\Gamma}{S}{\kity}$, the following hold:
\begin{enumerate}[\em(1)]
 \item\label{enum:comp:1} If in every $\Sigma$-model $I$ we have
  \[\STerms{\semc{\Gamma}{S}{I}}\neq\es,\] then there is a term $s$ with \[\oftype{\Sigma}{\Gamma}{s}{S}.\]
 \item\label{enum:comp:2} For all terms $\oftype{\Sigma}{\Gamma}{s}{S}$ and $\oftype{\Sigma}{\Gamma}{s'}{S}$, if $\semc{\Gamma}{s}{I}=\semc{\Gamma}{s'}{I}$ holds for all $\Sigma$-models $I$, then $\isequal{\Sigma}{\Gamma}{s}{s'}$.
\end{enumerate}
\end{theorem}
\begin{proof}
This follows immediately from Thm.~\ref{thm:mltt:modelexistence}, considering the term-generated model constructed there.
\myqed\end{proof}
 
Finally, observe that in the presence of extensional identity types, statement (\ref{enum:comp:1}) of Thm.~\ref{thm:mltt:complete} already implies statement (\ref{enum:comp:2}): For all well-formed terms $s$, $s'$ of type $S$, if $\semc{\Gamma}{s}{}=\semc{\Gamma}{s'}{}$ in all $\Sigma$-models, then $\semc{\Gamma}{\ident{s}{s'}}{}$ always has an element, and so there must be a term  $\oftype{\Sigma}{\Gamma}{t}{\ident{s}{s'}}$, whence $\isequal{\Sigma}{\Gamma}{s}{s'}$. An analogous result for types is more complicated and remains future work.

%% file: conclusion.tex
We have presented a concrete and intuitive semantics for MLTT in terms of indexed sets on posets. And we have shown soundness and completeness. Our semantics is essentially that proposed by Lawvere in \cite{lawvereadjoint} in the hyperdoctrine of posets, fibrations, and indexed sets on posets, but we have made particular choices for which the models are coherent. Our models use standard function spaces, and substitution has a very simple interpretation as composition. The same holds in the simply-typed case, which makes our models an interesting alternative to (non-standard) Henkin models. In both cases, we strengthen the existing completeness results by restricting the class of models.

We assume that the completeness result can still be strengthened somewhat further, e.g., to permit equality axioms between types. In addition, it is an open problem to find an elementary completeness proof, i.e., one that does not rely on topos-theoretical results.